\documentclass[a4paper,fleqn,usenatbib]{mnras}

\usepackage{newtxtext,newtxmath}

\usepackage[T1]{fontenc}
\usepackage{ae,aecompl}

\usepackage{graphicx}	
\usepackage{amsmath}	
\usepackage{xcolor}
\usepackage{natbib}
\usepackage{float}
\usepackage{hyperref}
\usepackage[normalem]{ulem}
\graphicspath{./Figures/}

\title[$\Lambda$CDM with baryons vs. MOND]{$\Lambda$CDM with baryons vs. MOND: the time evolution of the universal acceleration scale in the {\it Magneticum} simulations}

\author[A. C. Mayer et al.]{
Alexander C. Mayer,$^{1}$\thanks{E-mail: Al.Mayer@campus.lmu.de}
Adelheid F. Teklu,$^{1,2}$
Klaus Dolag$^{1,3}$
and Rhea-Silvia Remus$^{1}$
\\
$^{1}$Universit\"ats-Sternwarte, Fakult\"at für Physik, Ludwig-Maximilians-Universit\"at M\"unchen, Scheinerstr. 1, D-81679 München, Germany\\
$^{2}$Excellence Cluster Origins, Boltzmannstr. 2, D-85748 Garching, Germany\\
$^{3}$Max-Planck-Institute for Astrophysics, Karl-Schwarzschildstr. 1, D-85741 Garching, Germany\\
}

\date{Accepted XXX. Received YYY; in original form ZZZ}

\pubyear{2022}

\begin{document}
\label{firstpage}
\pagerange{\pageref{firstpage}--\pageref{lastpage}}
\maketitle

\begin{abstract}
MOdified Newtonian Dynamics (MOND) is an alternative to the standard Cold Dark Matter (CDM) paradigm which proposes an alteration of Newton's laws of motion at low accelerations, characterized by a universal acceleration scale $a_{0}$. It attempts to explain observations of galactic rotation curves and predicts a specific scaling relation of the baryonic and total acceleration in galaxies, referred to as the Rotational Acceleration Relation (RAR), which can be equivalently formulated as a Mass Discrepancy Acceleration Relation (MDAR). The appearance of these relations in observational data such as SPARC has lead to investigations into the existence of similar relations in cosmological simulations using the standard $\Lambda$CDM model. Here, we report the existence of an RAR and MDAR similar to that predicted by MOND in $\Lambda$CDM using a large sample of galaxies extracted from a cosmological, hydrodynamical simulation (\textit{Magneticum}).
Furthermore, by using galaxies in \textit{Magneticum} at different redshifts, a prediction for the evolution of the inferred acceleration parameter $a_{0}$ with cosmic time is derived by fitting a MOND force law to these galaxies. In \textit{Magneticum}, the best fit for $a_{0}$ is found to increase by a factor $\simeq$3 from redshift $z=0$ to  $z=2.3$. This offers a powerful test from cosmological simulations to distinguish between MOND and $\Lambda$CDM observationally.
\end{abstract}




\section{Introduction}

Dark matter forms an integral part of the $\Lambda$CDM model of cosmology. This form of matter, which does not interact with electromagnetic radiation, is thought to be largely responsible for the observed dynamics in individual galaxies as well as galaxy clusters. Additionally, it is assumed to be very important for the formation of structure in the Universe, and the dark matter content of the Universe also has large implications regarding cosmology. One particular observation for which dark matter offers an explanation is the high velocity at which stars and gas in the outer regions of galaxies move around the center of their host galaxy. However, from the distribution of visible matter (stars and gas) in these galaxies, Newtonian physics would predict a Keplerian ($\sim\frac{1}{\sqrt{r}}$) drop of the rotational speed contrary to the actual velocity observed at progressively greater radii from the galactic centre. Over the years, many alternative explanations have been proposed for these observations, one of the most well known of which is MOdified Newtonian Dynamics (MOND) (\cite{1983ApJ...270..365M}, \cite{1983ApJ...270..371M}, \cite{1983ApJ...270..384M}). MOND stipulates the existence of a universal acceleration scale, usually denoted $a_{0}$. Below this scale, significant differences between the usual Newtonian laws of motion and MOND are proposed. Explicitly, the relation between the Newtonian acceleration $a_{\mathrm{bar}}$ and the MOND acceleration $a_{\mathrm{tot}}$ is given by:
\begin{equation}
a_{\mathrm{bar}}=a_{\mathrm{tot}}\cdot\mu(\frac{a_{\mathrm{tot}}}{a_{0}}). \label{MOND_acc}
\end{equation}

The $\mu$ in this formula is the \textit{interpolation function}, which, while not entirely specified by MOND, must meet two requirements. In the limit of high accelerations, it has to reproduce Newton's second law, meaning $\mu(x)\rightarrow 1$ for $x$>>1. On the other hand, MOND must be consistent with the observation of flat rotation curves, which leads to the requirement that $\mu(x)\rightarrow x$ for $x$<<1.
Equation \eqref{MOND_acc} predicts that for a given value of the Newtonian acceleration from stars and gas in a galaxy, there is a specific value of the total acceleration that should be measured if MOND is indeed the explanation for the shape of galactic rotation curves. This relation between the baryonic acceleration (with Newtonian physics) $a_{\mathrm{bar}}$ and the total acceleration $a_{\mathrm{tot}}$ is referred to as the \textit{Rotational Acceleration Relation} (\textit{RAR}). An equivalent formulation is given by the \textit{Mass Discrepancy Acceleration Relation} (\textit{MDAR}):

For a spherically symmetric distribution of matter,
\begin{equation}
a(r)=\frac{GM(<r)}{r^{2}} \label{GMoverR}
\end{equation}
($M(<r)$ is the mass enclosed in a sphere with radius r), which can be used to reformulate equation \eqref{MOND_acc} to the MDAR:
\begin{equation}
\frac{M_{\mathrm{dyn}}}{M_{\mathrm{bar}}}=\frac{1}{\mu(\frac{a_{\mathrm{tot}}}{a_{0}})}. \label{MDAR}
\end{equation}

We will \textit{define} $M_{\mathrm{dyn}}$ and $M_{\mathrm{bar}}$ by relation \eqref{GMoverR} with $a=a_{\mathrm{tot}}$ and $a=a_{\mathrm{bar}}$, respectively (and they would only be identical to the actual enclosed mass in a perfectly spherical galaxy). The impact of dropping this simplification will be discussed in section \ref{check}.  \\
A common choice for $\mu$ is the ``simple interpolation function''
\begin{equation}
\mu (x)=\frac{x}{1+x}, \label{simplemu}
\end{equation}
yielding the MDAR-formula: 
\begin{equation}
\frac{M_{\mathrm{dyn}}}{M_{\mathrm{bar}}}(r)=1+\frac{a_{0}}{a_{\mathrm{tot}}(r)} \label{MDARsimple}
\end{equation}
For the RAR we use a different, but also commonly used, relation:
\begin{equation}
a_{\mathrm{tot}}=\frac{a_{\mathrm{bar}}}{1-e^{-\sqrt{\frac{a_{\mathrm{bar}}}{a_0}}}}
\label{eq:a1}
\end{equation}
as first proposed by \cite{McGaugh_2008}. 

If the form of the relation is fixed, the only free parameter left in this formulation of MOND is the value of $a_{0}$ as a characteristic acceleration scale. In this work, this $a_{0}$ is used mainly as a fitting parameter for data from \textit{Magneticum} and is allowed to vary over cosmic time, leading to a relation $a_{0}(z)$.

A consequence of the MOND force law \eqref{MOND_acc} is the Baryonic Tully Fisher Relation (BTFR), which links the asymptotic rotation speed of a galaxy to the baryonic mass of that galaxy (and not the total luminosity): 
\begin{equation}
v_{\mathrm{asymptotic}}^{4}=G a_{0} M_{\mathrm{bar}} \label{BTFR}
\end{equation}
Especially important to this relation is the exponent of exactly 4. 

In summary, MOND makes concrete predictions for relations between the distribution of baryonic matter in a galaxy and the dynamics of that galaxy. 
In a $\Lambda$CDM universe or simulation on the other hand, a tight relation between these two seems unlikely at first, as DM is supposed to dominate the dynamics of galaxies. 
However, many apparent problems with $\Lambda$CDM, as for example the formation of too large bulges of disk galaxies, the cusp/core problem or the too big to fail problem, have been resolved by the introduction of baryonic feedback processes such as outflows driven by Active Galactic Nuclei (AGN) and Supernovae \citep[see e.g.][]{Governato2010,Brook2011,Governato2012,Maccio2012,Brooks2014,Ogiya2014}. 
It is therefore not unreasonable to assume that such baryonic physics can also lead to similar relations as those predicted by MOND.

The predictions of MOND for the MDAR and RAR have been shown to 
reflect observations of galaxies in the \textit{SPARC} (``Spitzer Photometry \& Accurate Rotation Curves'')
sample (\cite{2016PhRvL.117t1101M}), which will be described in more detail below. But earlier work on simulations of individual galaxies has shown that such relations also naturally arise in $\Lambda$CDM simulations (\cite{2017MNRAS.471.1841N}, \cite{2019MNRAS.485.1886D}), which seems to indicate that a tight relation between baryonic mass distribution and galactic dynamics is expected in a $\Lambda$CDM universe. The possibility of a changing characteristic acceleration scale $a_{0}$ over cosmic time has also been investigated (\cite{2017ApJ...835L..17K}).

The question of the existence of similar acceleration relations in a statistically relevant sample of galaxies from the \textit{Magneticum} simulations is the main point of interest in this work. \textit{Magneticum} provides a large, representative sample of galaxies with different masses and different morphologies at several redshifts, making it well suited to obtain a $\Lambda$CDM prediction for the relation $a_{0}(z)$. 

This paper is structured as follows:
Firstly, we describe the \textit{Magneticum} simulations. Then, we describe the SPARC dataset, which serves as a point of comparison of results form \textit{Magneticum} to observations. Data from \textit{Magneticum} is then discussed in the context of predictions from MOND, in particular the BTFR, MDAR and RAR. Then, the dependence of the characteristic acceleration in \textit{Magneticum} as a function of redshift is focused on. The next step is the discussion of possible sources of error. After that, the discussion of the MDAR and RAR is extended to a larger class of galaxies, and characteristic quantities of the Magneticum galaxies are considered. Lastly, the results from \textit{Magneticum} are compared both to predictions by MOND as well as by other $\Lambda$CDM simulations. 

\section{The \textit{Magneticum} simulations and SPARC observations}

\begin{figure}
\includegraphics[width=0.46\textwidth]{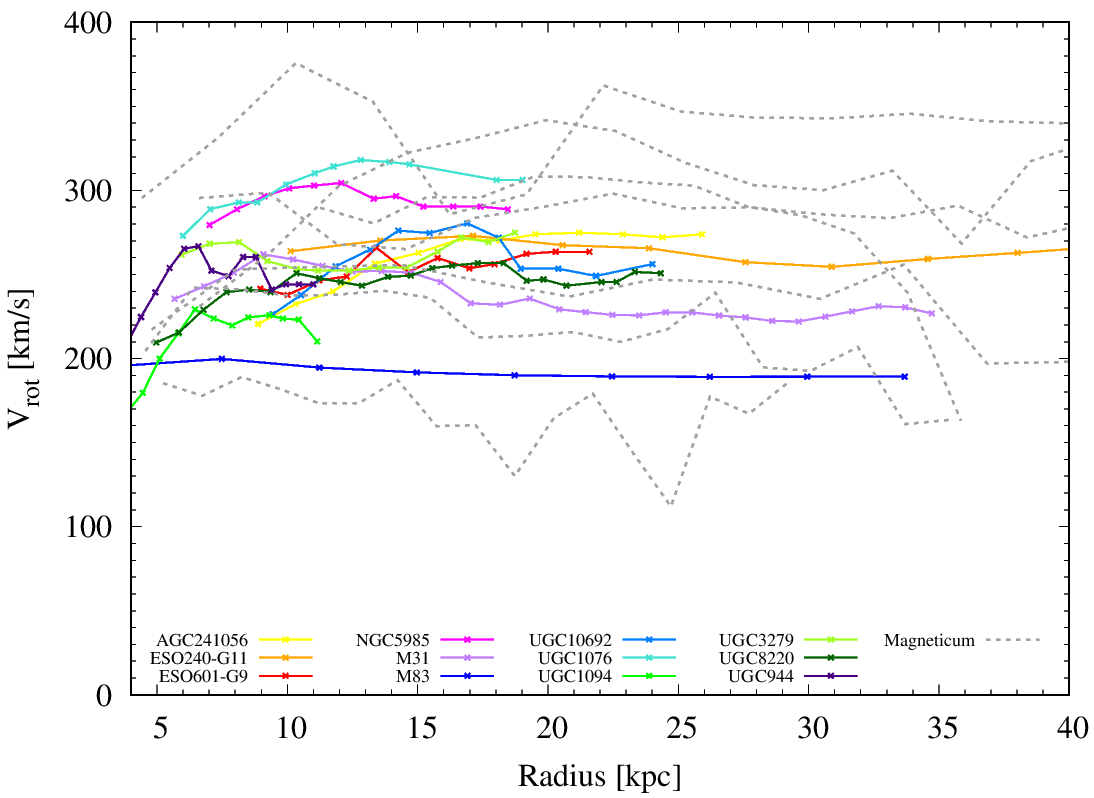}
\caption{Rotation curves of selected \textit{Magneticum} galaxies (gray dashed lines), computed from the actual tangential velocity of the according gas particles, compared to measured rotation curves from a set of observed spiral galaxies (see text for details). 
}
\label{rotcurveBTF136}
\end{figure}

\subsection{The \textit{Magneticum} simulations}
\label{Magneticum_simulation}

The \textit{Magneticum} simulations are a set of state-of-the-art, cosmological, hydrodynamical simulations of different cosmological volumes with different resolutions, performed with an improved developers' version \citep[see][for details on the numerical scheme]{Beck15} of the N-body/SPH code \textsc{Gadget-3}, which in turn is an updated version of the well-known open-source code \textsc{Gadget-2} \citep{2005MNRAS.364.1105S}.  
They follow a standard $\Lambda$CDM cosmology with parameters ($h$, $\Omega_{M}$, $\Omega_{\Lambda}$, $\Omega_{b}$, $\sigma_{8}$) set to
($0.704$, $0.272$, $0.728$, $0.0451$, $0.809$), adopting a WMAP 7 cosmology \citep{2011ApJS..192...18K}.

 These simulations follow a wide range of physical processes \citep[see][for a very detailed description]{Hirschmann14a,2015ApJ...812...29T} which are important for studying the formation of AGN, galaxies, and galaxy clusters.
In short, this includes radiative cooling and heating by a uniform evolving UV background, a multi-phase sub-grid model to trace star formation \citep{2003MNRAS.339..289S}, coupled to the treatment of stellar evolution and chemical enrichment processes properly modeling chemo-energetic feedback by SNIa, SNII and AGB onto the surrounding gas \citep{2007MNRAS.382.1050T}. Cooling is implemented following the metallicity-dependent formulation presented in \citet{2009MNRAS.393...99W} using cooling tables produced by the publicly available CLOUDY photo-ionization code \citep{1998PASP..110..761F}. In addition, black holes are represented by sink particles, which grow by accretions from the surrounding gas and thereby allow to model AGN feedback in different modes \citep[see][]{Hirschmann14a,Steinborn16}. Furthermore, it follows the thermal conduction similar to \citet{2004ApJ...606L..97D} but with 1/20th of the classical Spitzer value \citep{1962pfig.book.....S} motivated by full MHD simulations including an anisotropic treatment of thermal conduction \citep{2014arXiv1412.6533A}.

This allows to reproduce the properties of the large-scale, intra-galactic, and intra-cluster medium, among them the Sunyaev-Zeldovich signal induced by galaxy clusters, their pressure profiles, their baryon fraction and their intra-cluster light component \citep[see][]{Dolag16,Gupta17,2022A&A...663L...6A,Remus17a}. The simulated galaxies follow the observed mass-metalicity relations and the ICM shows the right composition of the different chemical species \citep{Dolag17,Kudritzki2021}. Also, basic properties of the black hole /  AGN population are reproduced, among them mass functions, luminosity functions as well as basic scaling relations between host galaxies and black hole properties \citep{Hirschmann14a,Steinborn16}.
Especially, detailed properties of galaxies of different morphologies can be recovered and well match observational findings. For example their angular momentum properties and the evolution of the stellar mass--angular momentum relation with redshift \citep{2015ApJ...812...29T,Teklu16}, stellar kinematics of early type galaxies \citep{Schulze18,Schulze20}, the size-mass relations and their evolution \citep[see e.g.][]{Remus16,Remus2017b}, global properties like the fundamental plane \citep{Remus16} or dark matter fractions \citep{Remus2017b}, in-situ and ex-situ fractions \citep{Remus21}, the baryon conversion efficiency \citep[see e.g.][]{Steinborn15,Teklu17}, as well as chemical properties \citep{Dolag17,Kudritzki2021} are in reasonable agreement with current observations.

\begin{figure}
\includegraphics[scale=0.5]{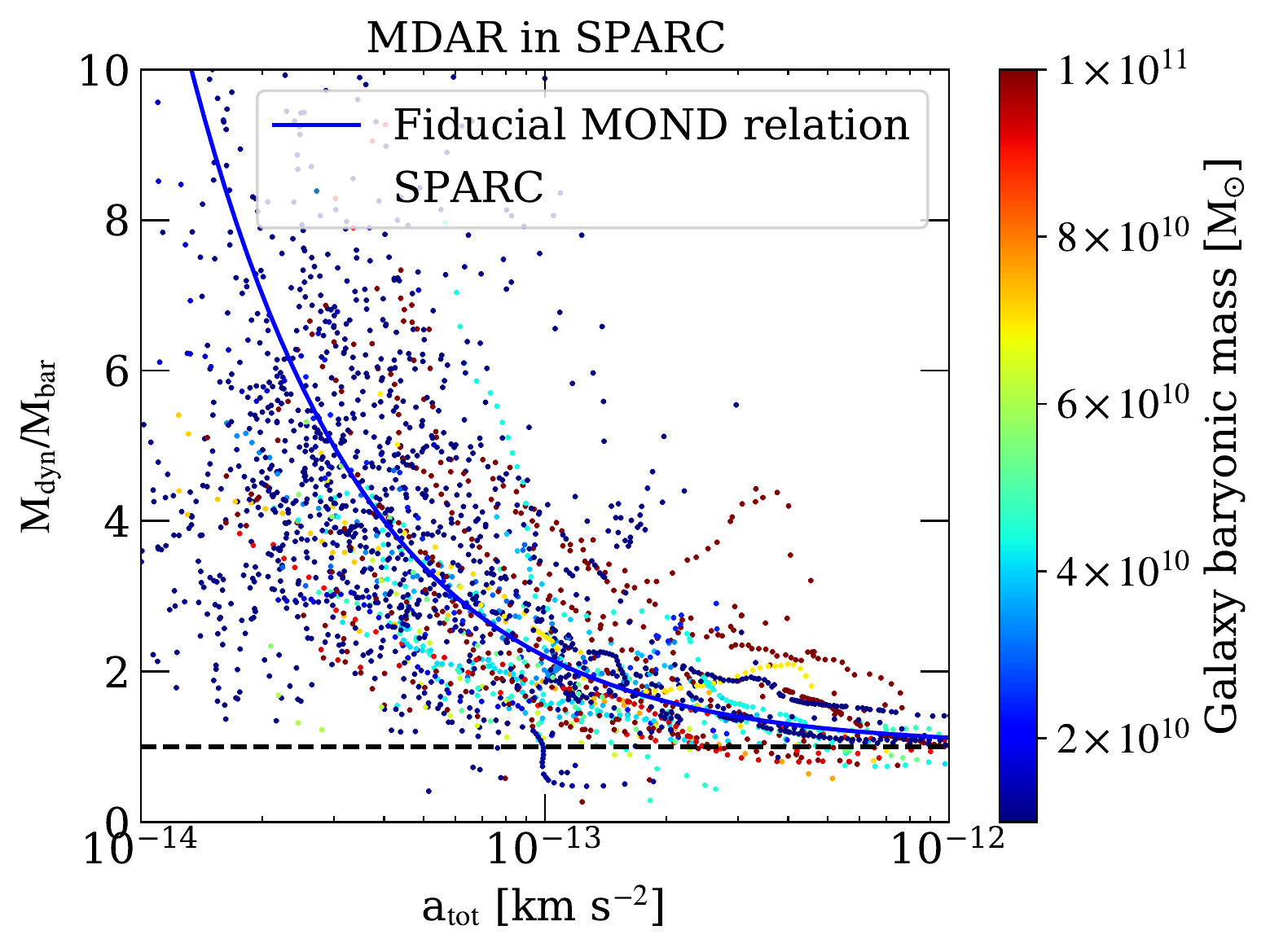}
\includegraphics[scale=0.5]{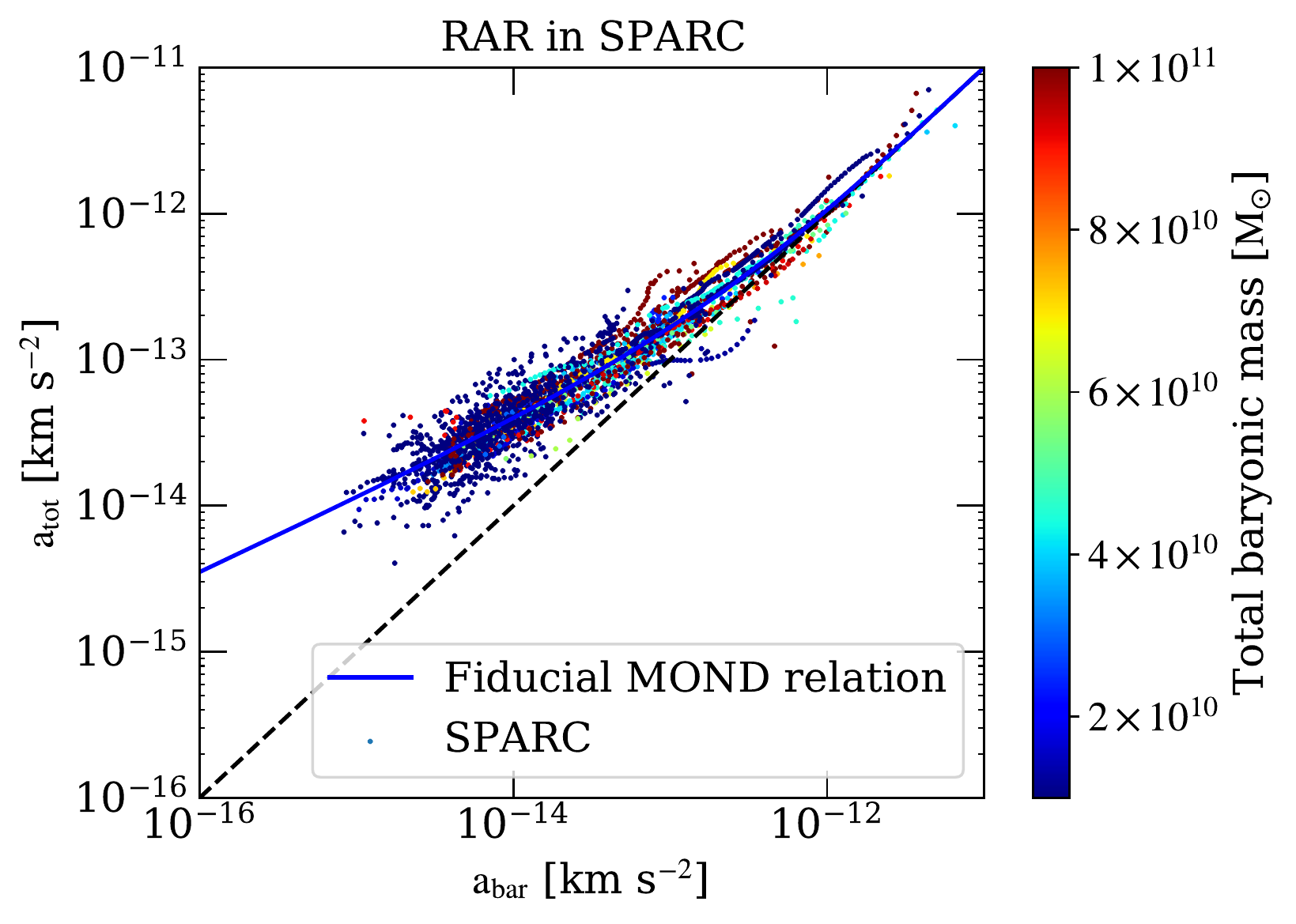}
\caption{Mass discrepancy acceleration relation (MDAR, upper) and Rotational acceleration relation (RAR, lower) obtained from the galaxies in the SPARC sample \citep{2016AJ....152..157L} compared with the MOND predictions using equations \eqref{MDARsimple} and \eqref{eq:a1}, respectively (displayed in blue, with $a_{0}=1.2\cdot10^{-13}\mathrm{km} \,\mathrm{s}^{-2}$). The black dashed lines are the newtonian relation. The colours encode the total baryonic mass of the galaxy as indicated in the colour bar.}
\label{SPARC}
\end{figure}

For this study we use the simulation \textit{Box4/uhr}, which covers a volume of (68 Mpc)$^{3}$, initially sampled with $2\cdot576^{3}$
particles (dark matter and gas), leading to a mass resolution of
$m_\mathrm{gas} = 7.3\cdot10^{6}M_\odot$ for the gas and $m_\mathrm{stars} \approx 1.8\cdot10^{6}M_\odot$ for stellar particles, with a gravitational softening of $0.7h^{-1}$kpc.
From this box we extract galaxies residing in haloes that have a virial mass higher than $5 \cdot 10^{11}M_{\odot}/h$ in order to ensure high enough resolution. 

We then classify the morphology of the galaxies in the simulation according to their position on the stellar-specific-angular-momentum--stellar-mass plane (see e.g. \cite{RF12}), i.e. the $b$-value \citep{2015ApJ...812...29T}, which is defined as 
\begin{equation}
b=\mathrm{log}_{10}(\frac{j}{\mathrm{kpc \, km \, s^{-1}}})-\frac{2}{3}\mathrm{log}_{10}(\frac{M}{M_{\odot}}),\label{bvalue}
\end{equation}
where $j$ and $M$ are the total specific angular momentum and mass of stars within $\frac{1}{10}r_{\mathrm{vir}}$ of the galaxy, respectively. 
It has been shown to be a good classification scheme that reproduces well observed relations, e.g. the size--mass relation and its evolution \citep{Remus2017b}, as well as the fundamental plane distributions \citep{Remus16}.

We define disk galaxies as galaxies that satisfy 
\begin{equation}
b(z)>-\frac{1}{2}\mathrm{log}_{10}(1+z)+b_{0}
\end{equation}
with $b_{0}$=-4.357 being the value at $z=0$.
The $z$-dependence is a consequence of the evolution of the critical overdensity $\Delta_{\mathrm{c}}$ with redshift, which is explained in detail by \cite{2015ApJ...815...97O}. 
The total numbers of disk and spheroidal galaxies obtained from the simulations at the different redshifts used in this paper are displayed in Figure \ref{galaxyfigure}.

Figure \ref{rotcurveBTF136} shows rotation curves of selected \textit{Magneticum} disk galaxies at $z=0.1$ (gray dashed lines) in comparison to observed spiral galaxies. The data for M83 were provided by Peter Kamphuis and Baerbel Koribalski and those for the other galaxies were taken from \cite{Yegorova2011} with different original sources as follows: they used H$\alpha$ data for UGC944, UGC1094, UGC3279 and UGC8220 from \cite{Vogt2004}, for UGC1076, UGC10692 and AGC241056 from \cite{Catinella2006}, for NGC5985 from \cite{Blais-Ouellette2004}, for ESO601-G9 from \cite{Persic1996}, and HI data for ESO240-G11 from \cite{Kregel2004} and for M31 from \cite{Corbelli2010}. 
The rotation curves for our galaxy sample are directly obtained from the averaged circular velocities of the individual cold gas particles. 
In order to ensure that only gas within the disk contributes to the rotation curve, we used particles within a height of ±3 kpc above the plane of the disk. 

\begin{figure*}
\includegraphics[scale=0.5]{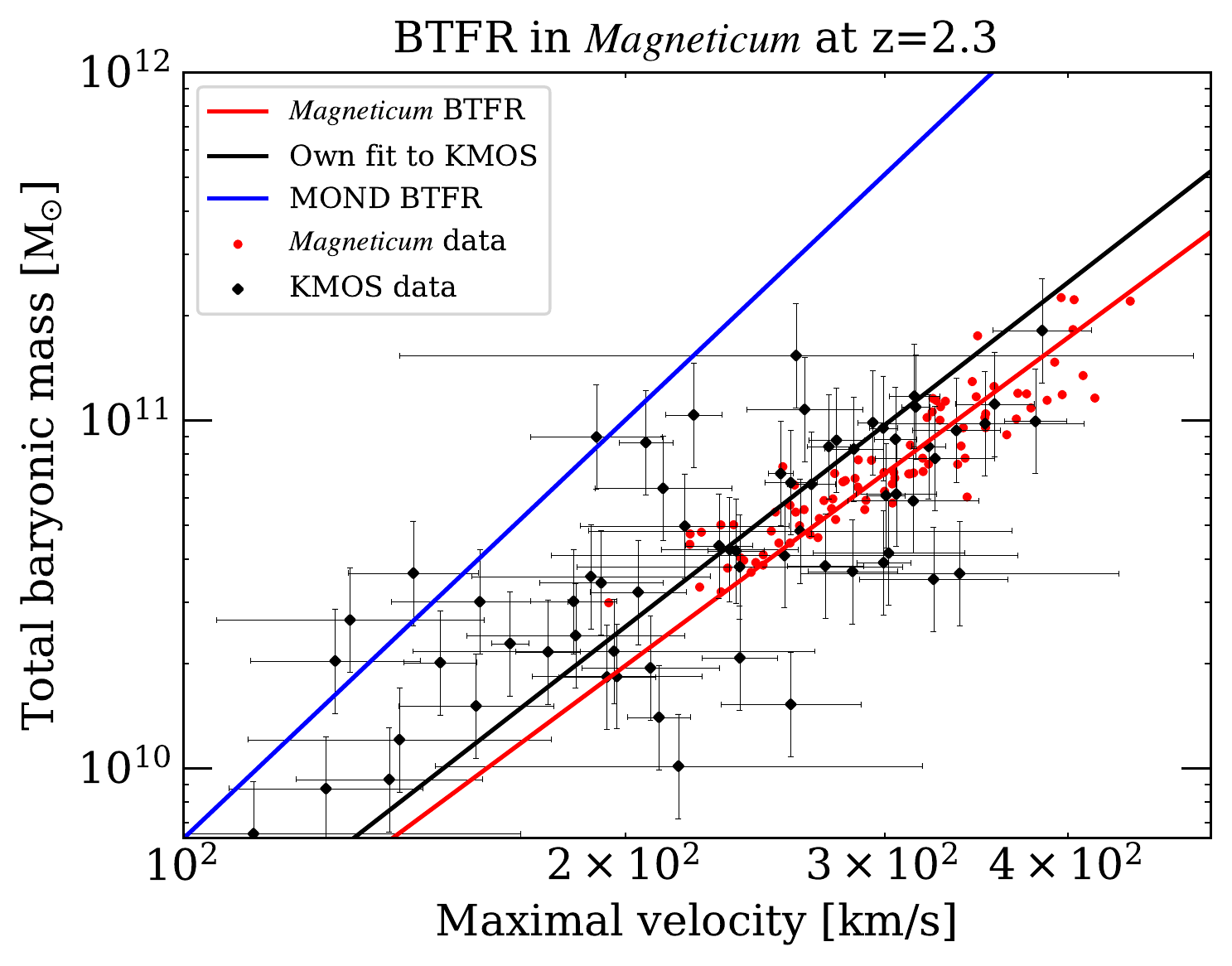}
\includegraphics[scale=0.5]{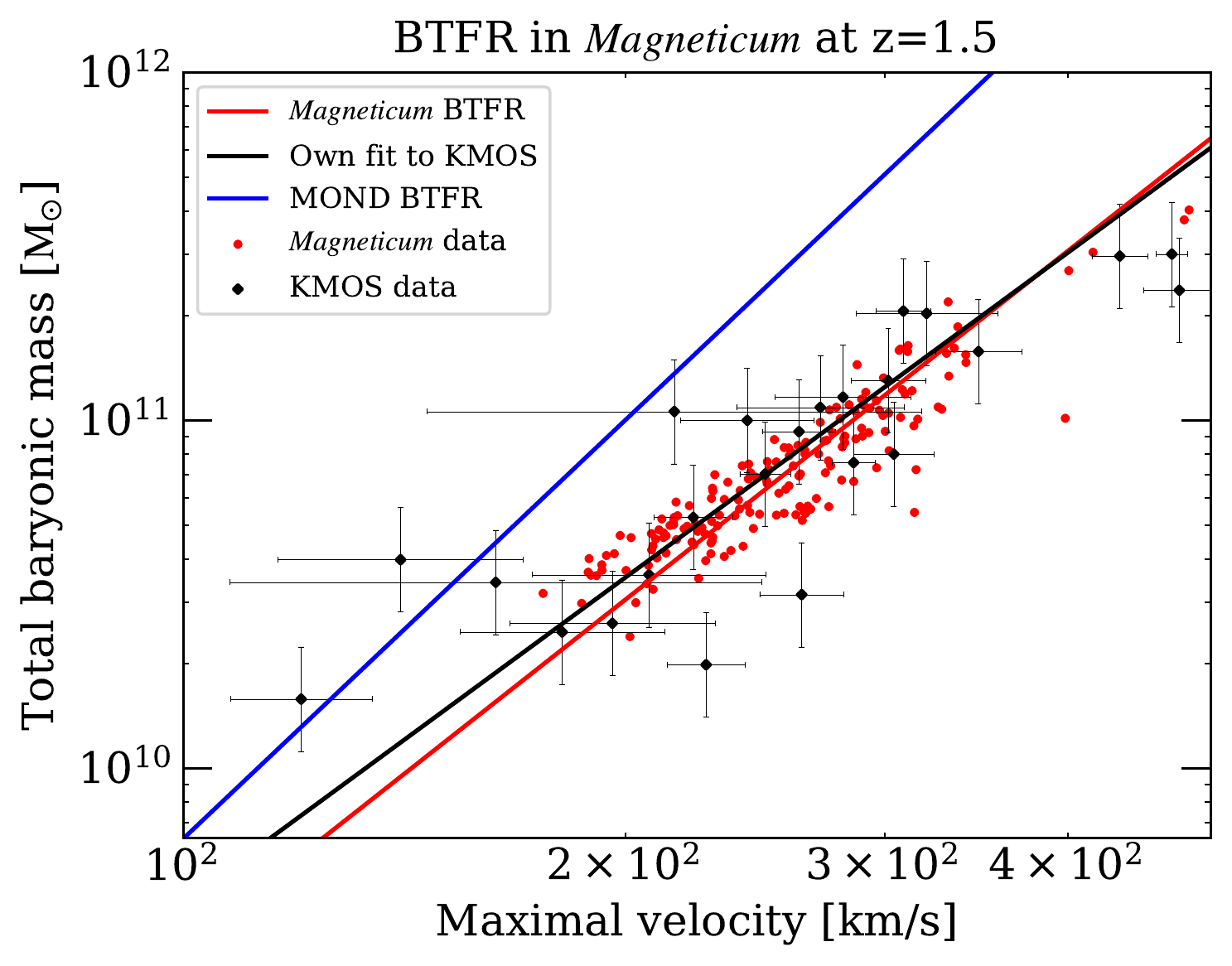}
\includegraphics[scale=0.5]{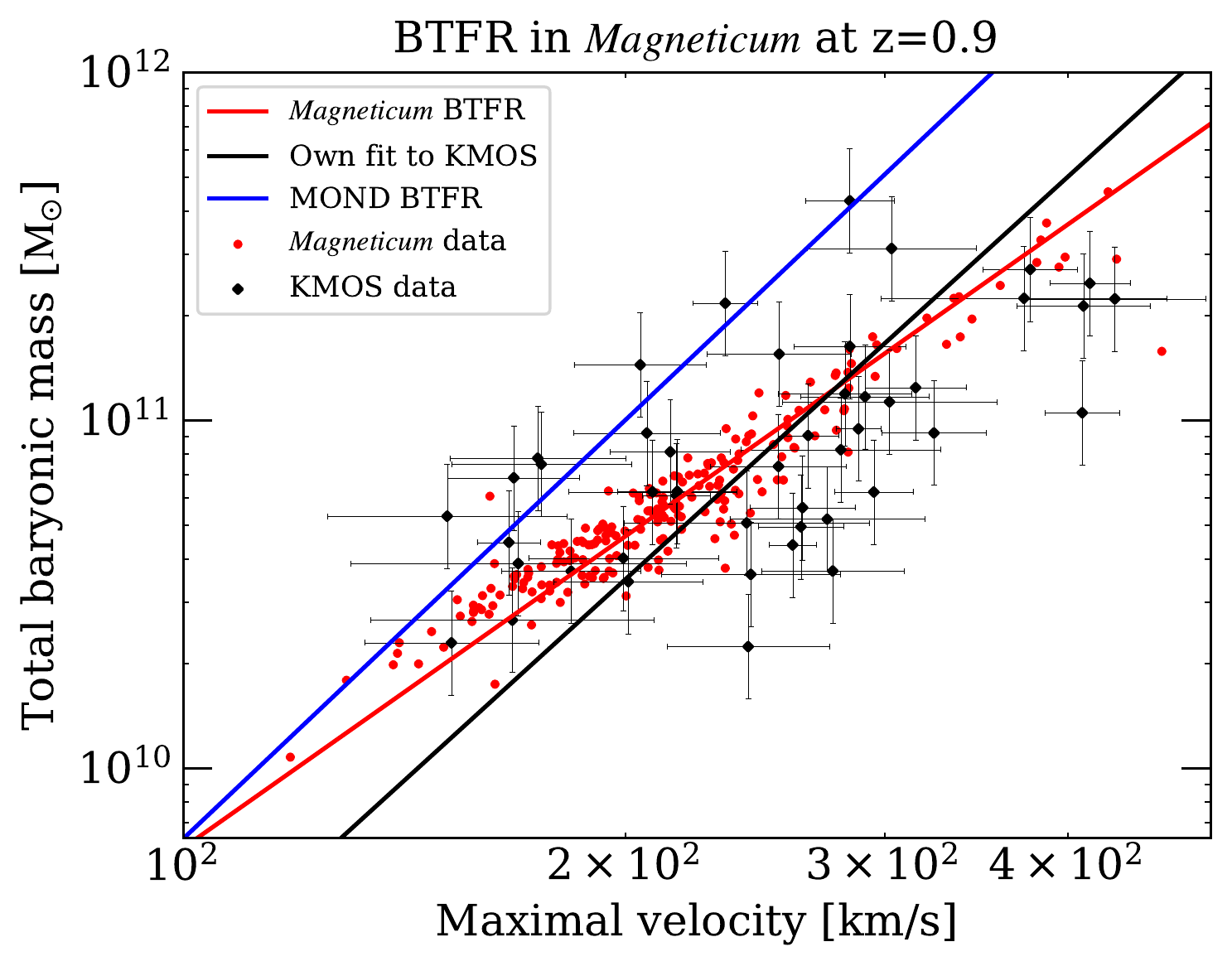}
\includegraphics[scale=0.5]{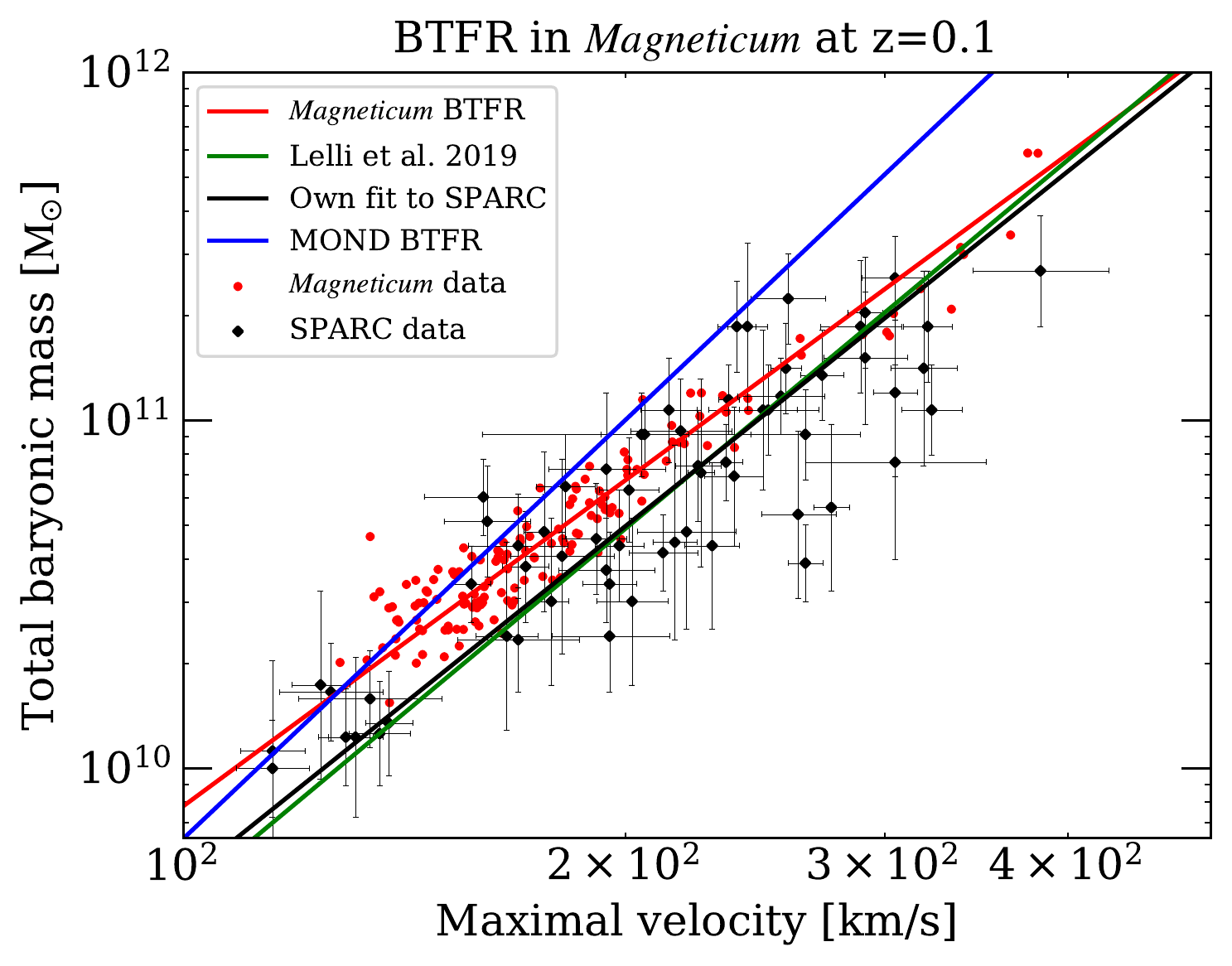}
\caption{Baryonic Tully Fisher realation (BTFR) at 4 different redshifts for the galaxies from the \textit{Magneticum} simulations (red), compared to observational data (\protect\cite{2017ApJ...842..121U}, \protect\cite{2019MNRAS.484.3267L}) displayed as black data points with error bars, as well as the MOND prediction from equation \eqref{BTFR} (blue). The total baryonic mass refers to the mass of all stars and gas within $\frac{1}{10}r_{\mathrm{vir}}$ of the galaxy, and maximal velocity refers to the estimated maximum of the rotational velocity calculated from the total enclosed mass (including dark matter). The fit values for the lines displayed are reported in table \ref{btfrtable}.}
\label{TFR_plot}
\end{figure*}

\begin{figure}
\includegraphics[scale=0.5]{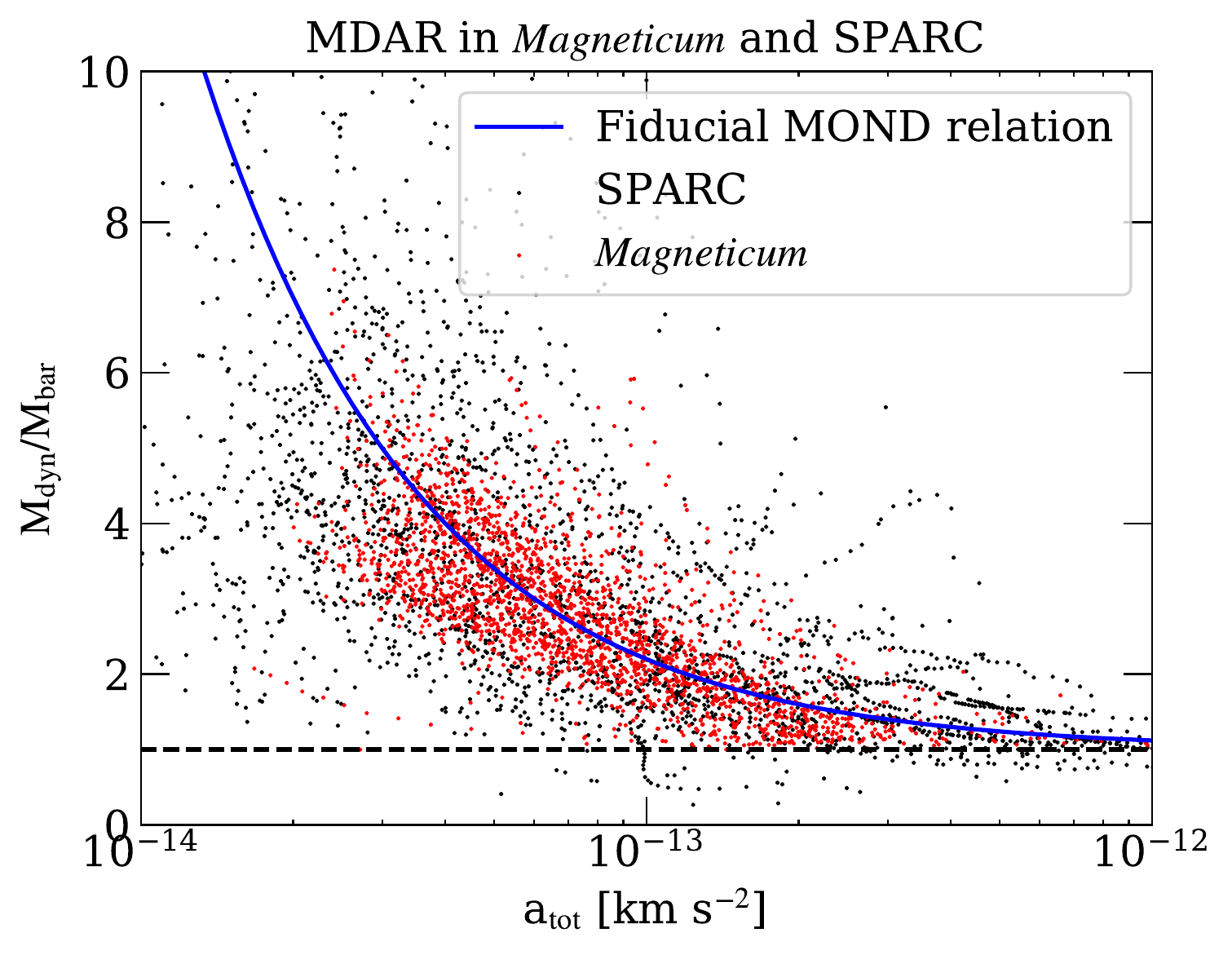}
\includegraphics[scale=0.5]{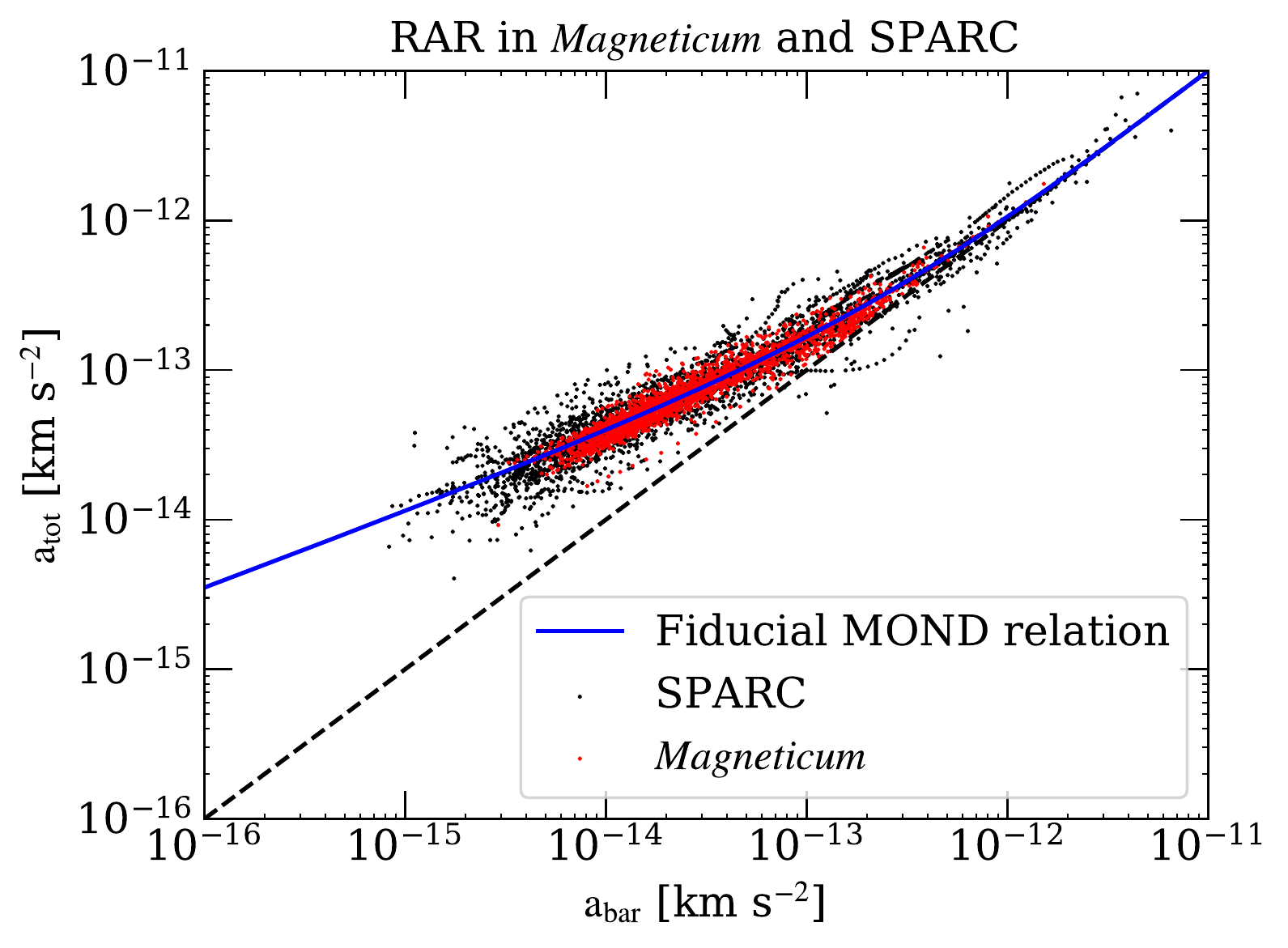}
\caption{Same as Fig. \ref{SPARC} but comparing galaxies in the SPARC sample (\protect\cite{2016AJ....152..157L}, black) with galaxies from the \textit{Magneticum} simulations  at $z=0.1$ (red). For the MOND predictions (blue) we used $a_{0}=1.2\cdot10^{-13}\mathrm{km} \,\mathrm{s}^{-2}$.}
\label{SPARC-Magneticum}
\end{figure}

\subsection{SPARC}
\label{sparc section}
The SPARC database (\cite{2016AJ....152..157L}) contains mass models and rotational velocities for 175 late type galaxies, where values for the mass models are split into three separate contributions from disk, bulge and gas. The data are stated with an assumed Mass-to-Light Ratio (MLR) of 1, which is a commonly used value, see e.g. \cite{2015ApJ...800..120Z}. Following earlier work (\cite{2016PhRvL.117t1101M}, \cite{2019MNRAS.485.1886D}), we assume an MLR of 0.7 for the bulge and 0.5  for the disk. While this is expected to vary from galaxy to galaxy, it has been argued (\cite{2016PhRvL.117t1101M}) that using the same values for all galaxies actually introduces the \textit{least} amount of assumptions and therefore bias into the results. However, the MLR in SPARC has also been investigated explicitly, see \cite{Schombert_2018} and \cite{Li_2018}. Both find good agreement of the average MLR with the values used here, although there is indeed the possibility of large deviations from galaxy to galaxy. \\
Additionally, we also excluded the same data points of poor quality as \cite{2016PhRvL.117t1101M}. This includes all measurements from 10 galaxies which are rejected due to low inclination ($i$<30) and from 12 where their asymmetric rotation curves are unlikely to trace the gravitational potential. Furthermore, individual data points from the remaining galaxies where the relative error of the observed velocity is larger than 10\% were excluded as well. 

In Fig. \ref{SPARC}, the data from SPARC clearly show an MDAR (upper panel)/RAR (lower panel) of the form predicted by MOND, as discussed by \cite{2016PhRvL.117t1101M}. This is true within the scatter of the individual galaxy tracks as well as for the mean of the sample. We note that there is no trend visible with the total baryonic mass (color-code). While \cite{2017ApJ...836..152L} argue that there is little or even no intrinsic scatter in the RAR, \cite{2018NatAs...2..668R} and \cite{2020MNRAS.494.2875M} conclude (using SPARC data among others) that there is a very low probability for a common acceleration scale throughout all galaxies.\\
The values for the total baryonic mass are discussed by \citet{2019MNRAS.484.3267L} and can be found on the SPARC website (http://astroweb.cwru.edu/SPARC/).

\begin{figure*}
\includegraphics[scale=1.2]{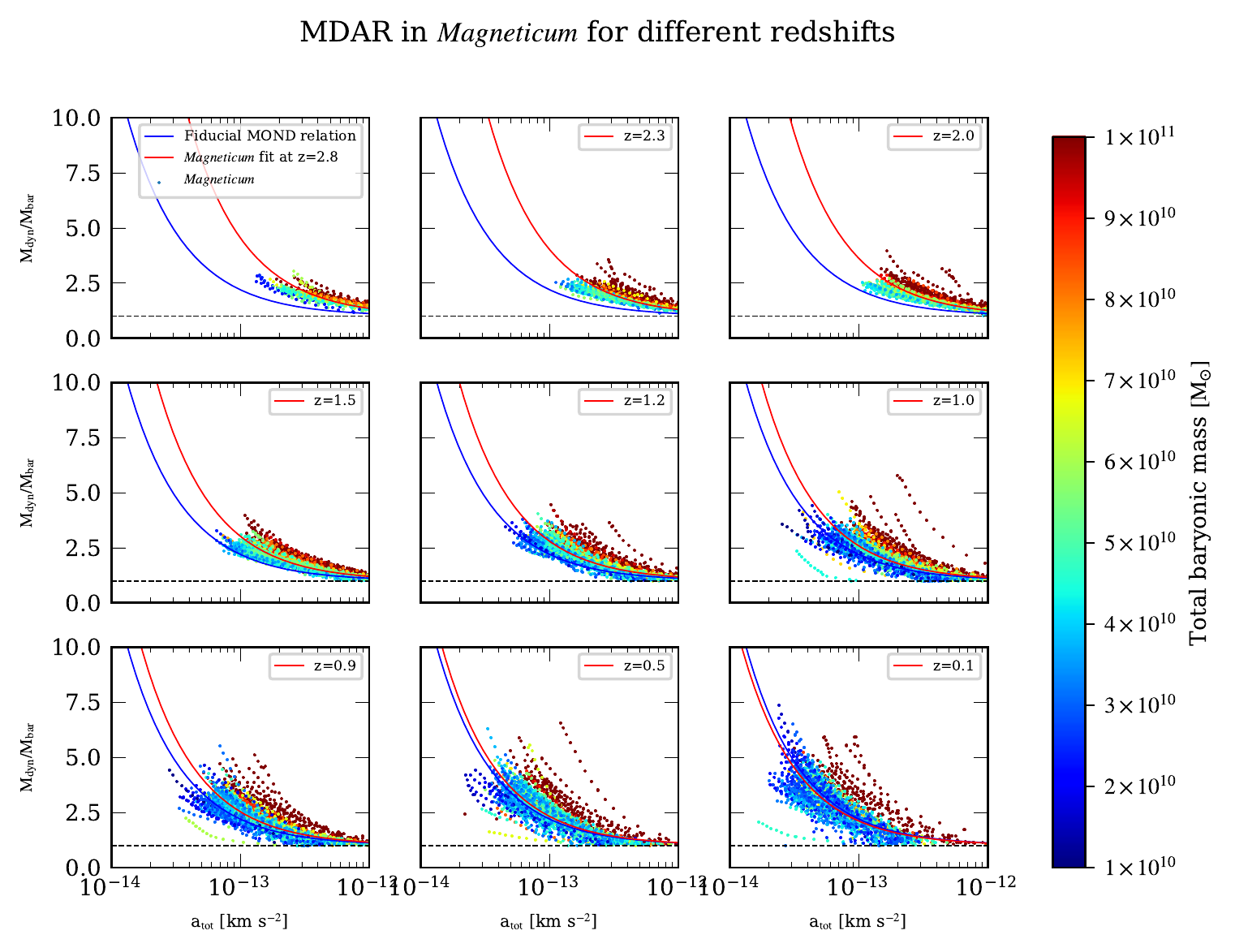}
\caption{Time evolution of the MDAR in \textit{Magneticum} from $z\approx3$ to $z\approx0$, as indicated in the panels. Points are from \textit{Magneticum} galaxies from each particular redshift and their colour indicates the total baryonic mass of the galaxy from which a given data point is taken. The function represented by the blue curve is computed from equation \eqref{MDARsimple} with a fiducial value of $a_{0}=1.2\cdot 10^{-13}\mathrm{km} \, \mathrm{s}^{-2}$, while the function shown as the red curve has the same form, but uses $a_{0}$ as a fitting parameter. A comparison between the results from the MDAR and the RAR relations at different redshifts can be found in the Appendix.}
\label{MDAR_z}
\end{figure*}

\section{Results from \textit{Magneticum}}

\subsection{TFR/BTFR in \textit{Magneticum}}
\label{BTFR_section}
In Figure \ref{TFR_plot}, the BTFR in \textit{Magneticum} is displayed for different redshifts, compared to observational data from two different sources. The SPARC dataset, used for the lowest redshift value, was described in \ref{sparc section}. For higher values of redshift, we utilize data from \cite{2017ApJ...842..121U} based on the KMOS$^{\mathrm{3D}}$-survey (see \cite{2015ApJ...799..209W} for the survey itself). The observations are in the range 0.6 < $z$ < 2.6, which, following \cite{2017ApJ...842..121U}, we divide into three bins with average values $z\approx$ 0.9, 1.5, 2.3 . The velocity data is given as the maximal modelled circular velocity, which includes a correction from gas pressure to the observed rotational velocity. We cannot apply a similar procedure to the  \textit{Magneticum} data and simply assume this will not significantly affect the results. \\
A similar range of masses and velocities is covered by the simulated and observational galaxies, although at high redshift \textit{Magneticum} does not resolve galaxies with masses as low as the observations display. The fits were performed using the \textit{scipy.odr} module, without taking the errorbars into account. \\
As is discussed among others by \cite{2018MNRAS.474.4366P}, \cite{2019MNRAS.484.3267L} and \cite{2021MNRAS.503.4147P}, the precise definition of the velocity used for the relation has a large impact on the resulting slopes. \cite{2019MNRAS.484.3267L} identify the asymptotic velocity as the one with the lowest scatter, and due to the fact that it is measured in the region of lowest baryonic acceleration, the MOND BTFR prediction of equation \eqref{BTFR} also most clearly applies there. We use $v_{max}$ instead of the asymptotic velocity in order to have consistent data between KMOS and SPARC, although the latter offers a number of relevant options. Using the asymptotic velocity results in larger exponents from both $Magneticum$ and SPARC.

\begin{table}
\caption{Resulting exponents from the power law fit to the BTFR for galaxies in {\it Magneticum} and observations at different redshifts, in contrast to the MOND prediction of 4.}
\begin{tabular}{ c | c | c | c | c }
\hline\hline
  & \multicolumn{2}{c}{number of galaxies} & \multicolumn{2}{c}{fit} \\
z & \textit{Magneticum} & Observation & \textit{Magneticum} & Observation \\
\hline
2.3 & 91 &65 & 3.13 & 3.29\\
1.5 & 169& 24 & 3.33 & 3.10\\
0.9 & 210 & 46 & 2.98 & 3.84\\
0.1 & 151 & 63 & 3.11 & 3.39\\
\hline
\end{tabular} \\
\label{btfrtable}
\end{table}
Clearly, the exponent of the BTFR in \textit{Magneticum} is inconsistent with the prediction by MOND at all redshifts larger than $z=0.1$. The \textit{Magneticum} galaxies show a smaller exponent than the data at the two values with lowest redshift as well as the largest one, substantially lower than the MOND prediction, see table \ref{btfrtable}. Except for at $z\approx 0.9$, the differences are not substantial, however. The \textit{Magneticum} exponent is actually larger than the observed one at $z\approx 1.5$. The observational data is sparse at this $z$-value, however. With accurate data for the total masses of galaxies, this difference could be used to clearly differentiate between $\Lambda$CDM and MOND. However, the accuracy of the BTFR inferred from observations is to a large part dependent on the MLR that is used to calculate the total baryonic mass of the galaxy, which leads to significant uncertainties in the BTFR. The importance of the MLR is stressed by \cite{2018MNRAS.474.4366P}, where the authors obtained $2.99\pm 0.22$ as the value for the exponent at $z=0.1$. On the other hand, \cite{2019MNRAS.484.3267L} calculate $3.85\pm 0.09$ from the galaxies in the SPARC sample at $z=0.1$. Both these studies utilized the asymptotic velocity.

\subsection{\textit{Magneticum} MDAR/RAR}
For all disk galaxies, we calculate the enclosed baryonic and total mass at 15 values of the radius, evenly spaced between $\frac{1}{200}$ and $\frac{1}{10}r_{\mathrm{vir}}$ of each galaxy. The minimum was chosen so small that it essentially represents a limit of the total to baryonic mass-fraction as one approaches the centre of the galaxy, while still being large enough such that some particles are always present within a sphere of that radius around the centre. The upper value on the other hand represents the outskirts of the galaxy, as there is very little baryonic matter belonging to the galaxy even further out. From these values for the masses, values for the ratio $\frac{M_{\mathrm{dyn}}}{M_{\mathrm{bar}}}(r)$ as well as $a_{\mathrm{tot}}(r)$ and $a_{\mathrm{bar}}(r)$ (using equation \eqref{GMoverR})  were calculated to obtain the \textit{Magneticum} MDAR and RAR, respectively, shown in Figure \ref{SPARC-Magneticum} for $z\approx 0$ in comparison to SPARC data. This neglects the contribution of turbulent motions onto the apparent rotation velocity of the gas \citep[see][for an example of such an effect]{2018ApJ...854L..28T} and assumes that this reflects what is done when fitting models to the observations to obtain dynamical masses and therefore shows a smaller scatter. So using the mass distributions directly can actually be seen as more truthful as the $\Lambda$CDM prediction. That is because if $\Lambda$CDM actually shows MOND-phenomenology, this must necessarily be a consequence of the mass distributions of galaxies in a $\Lambda$CDM universe. Note that the use of mass distributions with the approximation \eqref{GMoverR} prevents ratios $\frac{M_{\mathrm{dyn}}}{M_{\mathrm{bar}}}<1$ and $\frac{a_{\mathrm{tot}}}{a_{\mathrm{bar}}}<1$ from occurring. If the total acceleration is measured from the actual gas, such situations do arise, as can be seen in Figure \ref{SPARC} and \ref{SPARC-Magneticum} for SPARC.

The \textit{Magneticum}-MDAR at different redshifts is displayed in Figure \ref{MDAR_z}. At low redshifts, the galaxies extracted from \textit{Magneticum} fall onto the relation described by equation \eqref{MDARsimple}. Importantly, the very existence of any clear relation between baryonic and total acceleration in a simulation using $\Lambda$CDM is noteworthy, as there is no explicit implementation of any acceleration scale in such a simulation. Indeed it has been proposed (\cite{2015CaJPh..93..250M}) that the lack of such a relation in simulations with DM is both evident and sufficient to discard $\Lambda$CDM. However, \textit{Magneticum} clearly shows a relation of this kind at all redshifts, with a value of $a_{0}= (1.12\pm0.04)\cdot 10^{-13} \mathrm{km} \, \mathrm{s}^{-2}$ at $z=0.1$ for the MDAR and $a_{0}= (1.14\pm0.04)\cdot 10^{-13} \mathrm{km} \, \mathrm{s}^{-2}$ from the RAR (the average values and errors were calculated via a bootstrapping procedure). These results are consistent with $a_{0}=(1.20 \pm 0.02 \,\mathrm{(random)}\, \pm 0.24 \,\mathrm{(systematic)}\,) \cdot 10^{-13} \pm \mathrm{km} \, \mathrm{s}^{-2}$ found from SPARC (\cite{2017ApJ...836..152L}), which is unsurprising in light of the close agreement in Figure \ref{SPARC-Magneticum}.  \\

Although the individual scatter of the simulated galaxies around the mean relation is comparable between the simulations and the observations, the simulations show a clear,
general trend, where galaxies with higher total baryonic mass tend to show higher mass discrepancies than those with lower total baryonic mass. This trend can not be seen in the SPARC data as displayed in Figure \ref{SPARC}, where the observed galaxy by galaxy scatter is much larger and prevents a direct comparison of this predicted trend by the simulations.

\begin{figure*}
\includegraphics[scale=0.55]{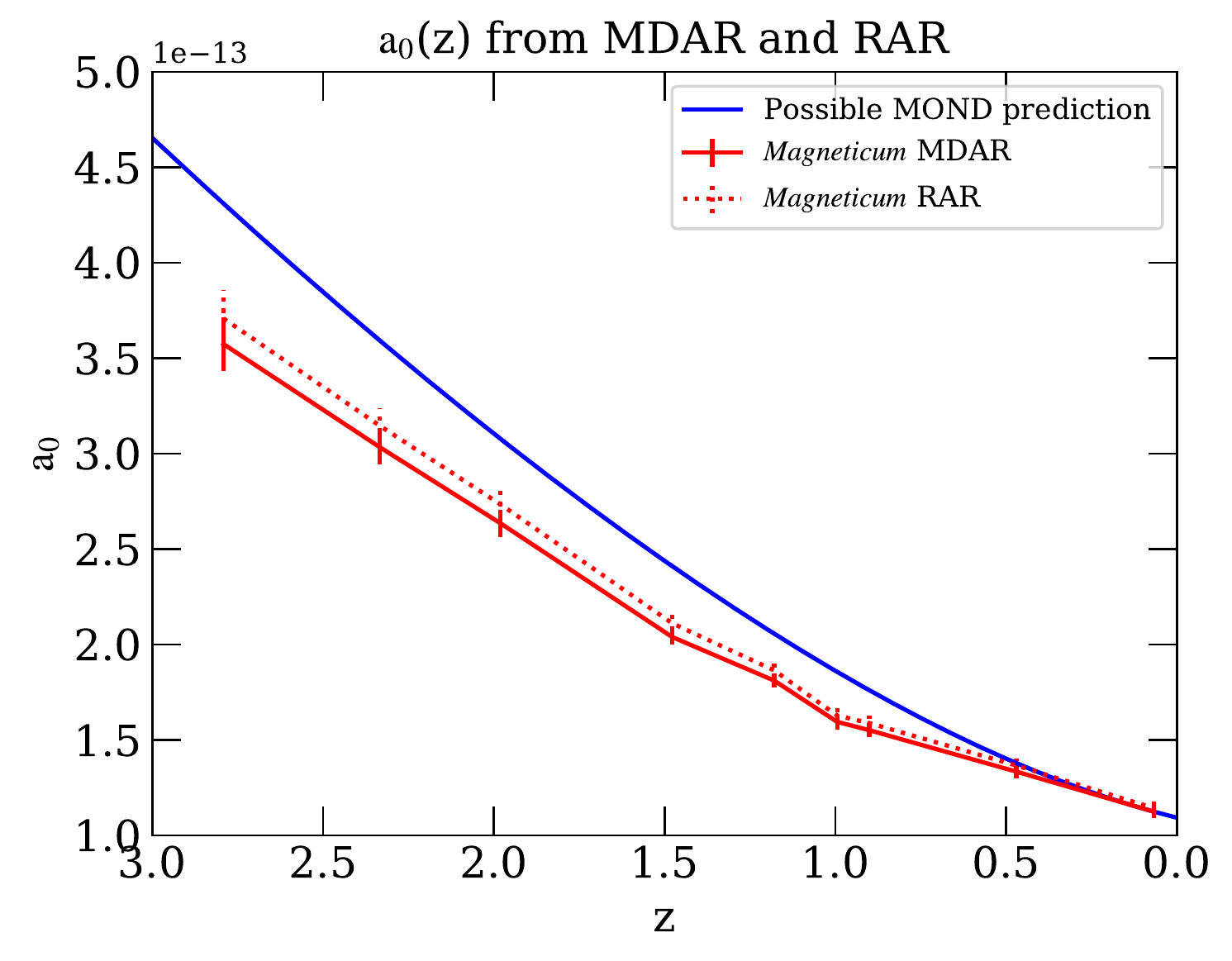}
\includegraphics[scale=0.55]{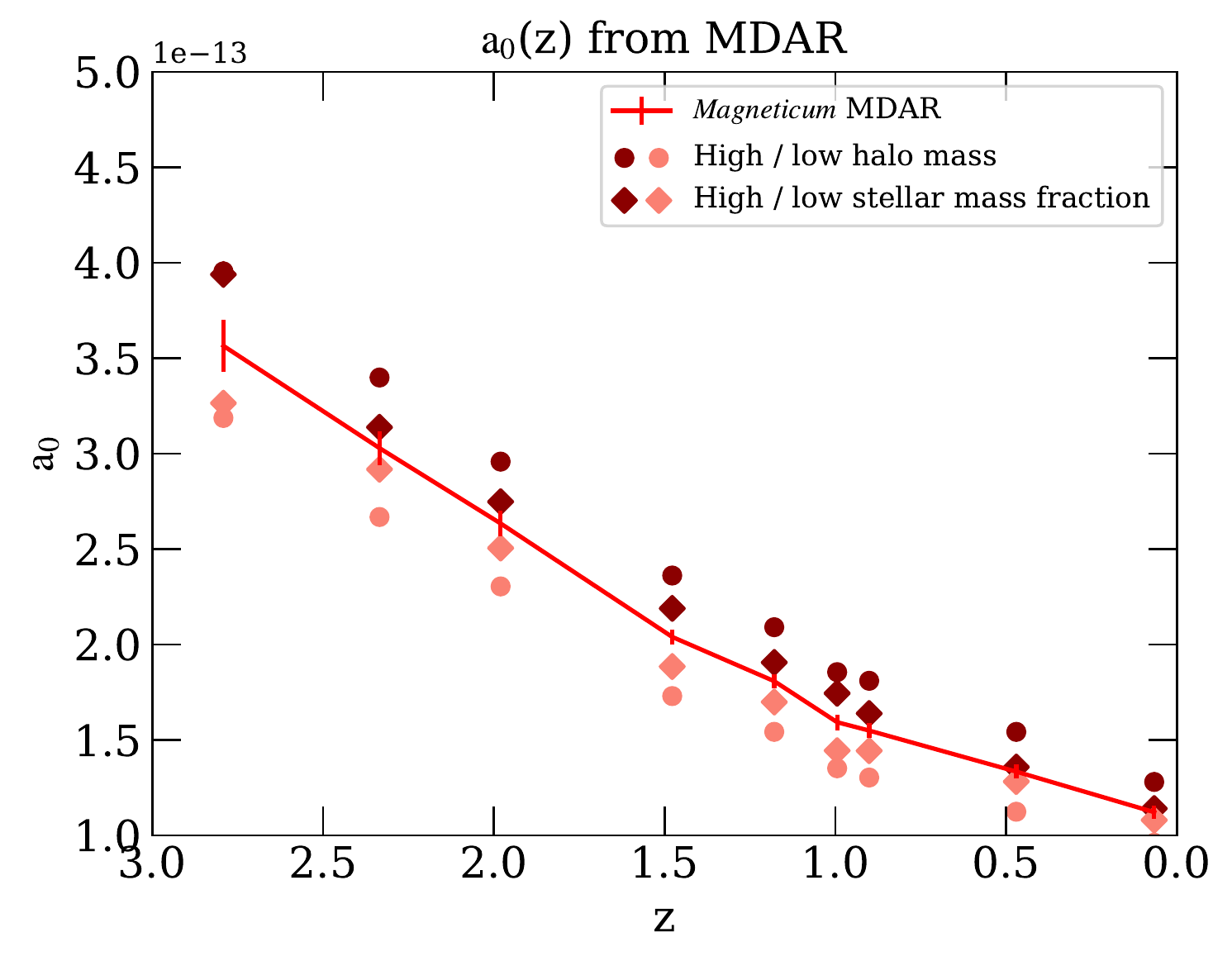}
\caption{Left: Best-fitting acceleration parameter $a_{0}$ to \textit{Magneticum} galaxies as a function of redshift (red lines), showing both values from the MDAR and RAR (solid and dotted line, respectively). The blue curve is obtained by evaluating equation \eqref{MONDnaive} (using the $Magneticum$ cosmological parameters), such that the value of $a_0$ agrees at the lowest redshift value. The decline with decreasing redshift can be clearly seen in all curves. Right: Effect of dividing the full sample into two sub-samples of equal size by halo mass (circles) or stellar mass fraction (diamonds) on the evolution $a_0 (z)$, for readability only showing the MDAR.}
\label{a0(z)}
\end{figure*}

\subsection{$a_{0}(z)$ in \textit{Magneticum}}

While these similarities at $z\approx 0$ could reflect the fact that both hydrodynamical simulations within the $\Lambda$CDM framework as well as MOND (by construction) do match the apparent matter distribution within galaxies, there is no a priori reason why this should still be the case at higher redshift. Interestingly, also at higher redshifts, galaxies extracted from the \textit{Magneticum} simulations can still be fitted remarkably well by a function of the form  of equation \eqref{MDARsimple}, as shown in Figure \ref{MDAR_z}, showing the MDAR from the simulations up to a redshift of $z\approx3$.

At higher redshifts, the best-fitting value for $a_{0}$ predicted by the \textit{Magneticum} simulation data obtained by fitting a function of the form of equation \eqref{MDARsimple} changes substantially. Here, Figure \ref{MDAR_z} shows a distinct decline of the fit-parameter $a_{0}$ with decreasing redshift. The relation between $a_{0}$ and $z$ is displayed in Figure \ref{a0(z)}, fitted to both the MDAR and RAR (in the latter case using the formula \eqref{eq:a1}). The left panel also includes a possible MOND prediction, to be discussed in section \ref{mond_pred}. The decrease of $a_{0}$ with cosmic time corresponds to a decrease of mass/acceleration discrepancies at lower redshifts, i.e. galaxies become more dominated by baryonic matter as time progresses and therefore differ substantially from simple expectations within the MOND framework. This is mainly a consequence of the cooling of gas, which leads to the accumulation of baryonic matter over time. Indeed, \cite{2017ApJ...835L..17K} argued that baryonic feedback processes such as supernovae play little role in the change of the characteristic acceleration scale over time, as the evolution of this scale over time is also present in simulations without such feedback. \\
The right panel of Figure \ref{a0(z)} compares the results for the MDAR from the full sample of disk galaxies (as before in the left panel) and the values obtained from dividing the sample into two halves according to halo mass or stellar mass fraction (both referring to mass within the virial radius). The division is performed at each  fixed redshift value individually. More massive haloes consistently yield higher values of $a_0$, which is unsurprising in light of the findings in the previous section of higher mass discrepancies in galaxies with higher total \textit{baryonic} mass (there within $\frac{1}{10} r_{\mathrm{vir}}$, however). High and low mass halos show basically the same evolution as the full sample, but shifted up or down, respectively. This is somewhat less stable if the stellar mass fraction is used as a divider, where high values correlate with larger $a_0$ - a separate trend from the above, as more massive haloes actually tend to have lower stellar mass fractions. One would intuitively expect the opposite, higher $a_0$ from lower fractions. It seems that higher acceleration discrepancies are generally found if the centre (here within $\frac{1}{10} r_{\mathrm{vir}}$) of a galaxy halo has a comparatively high total acceleration (which can both be the case for high halo mass and stellar mass fraction).

\section{Discussion}

\subsection{Comparison with results from potential theory}
\label{check}
The intrinsic scatter of the acceleration relations could be a useful tool in distinguishing whether MOND or DM provides the `missing acceleration'. If total and baryonic accelerations could be measured with perfect accuracy, then MOND would not allow for any scatter in the resulting relation. In practice, a central obstacle for such perfect measurements are again uncertainties in the MLR, see e.g. \cite{2016AJ....152..157L}. This is not a problem in simulations, but the $\Lambda$CDM prediction is a priori less clear. Intrinsic scatter in the MDAR/RAR would not be surprising, however, as this is just equivalent to a variation of the halos to stellar mass relation between different galaxies.

\begin{figure}
\includegraphics[scale=0.25]{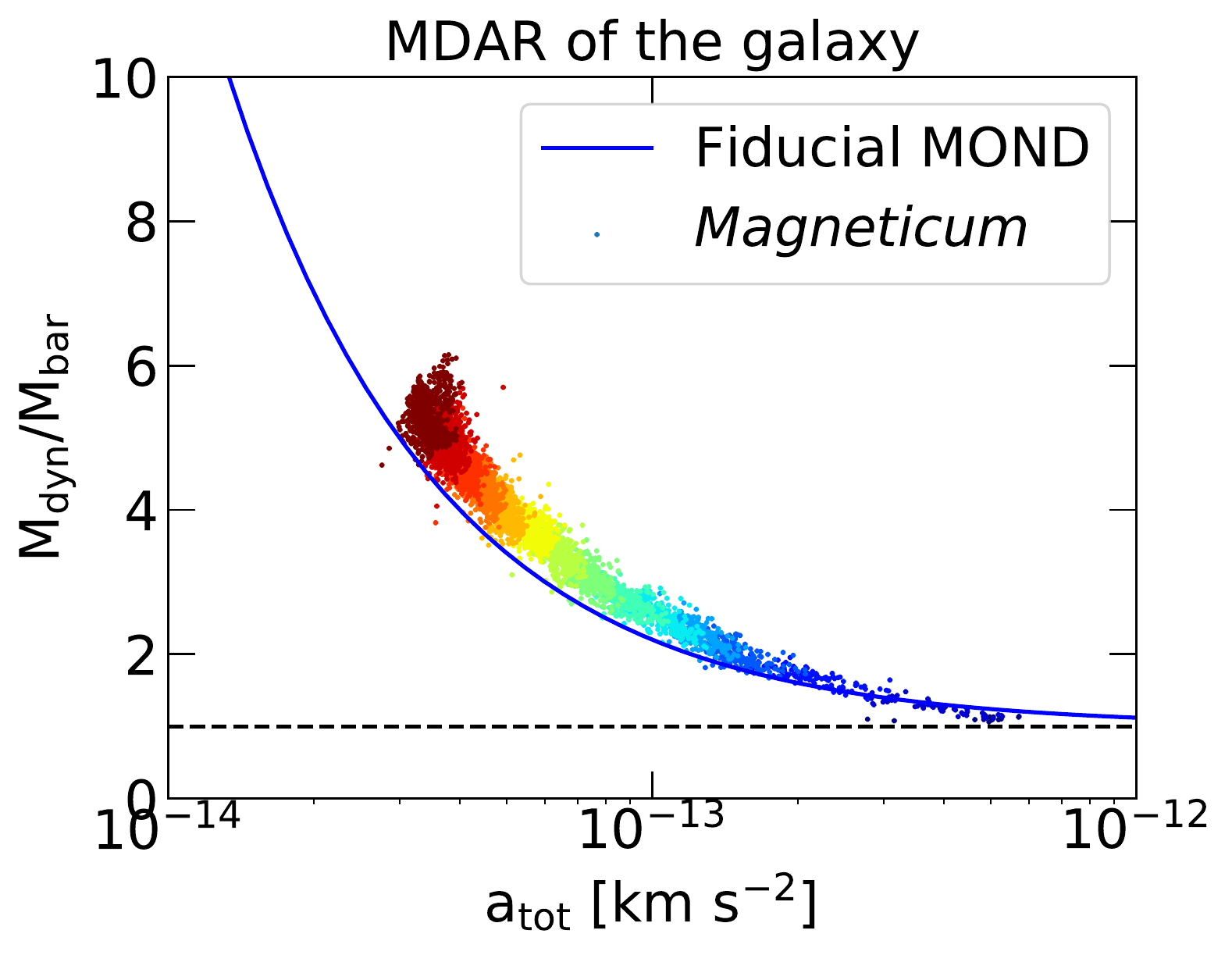}
\includegraphics[scale=0.25]{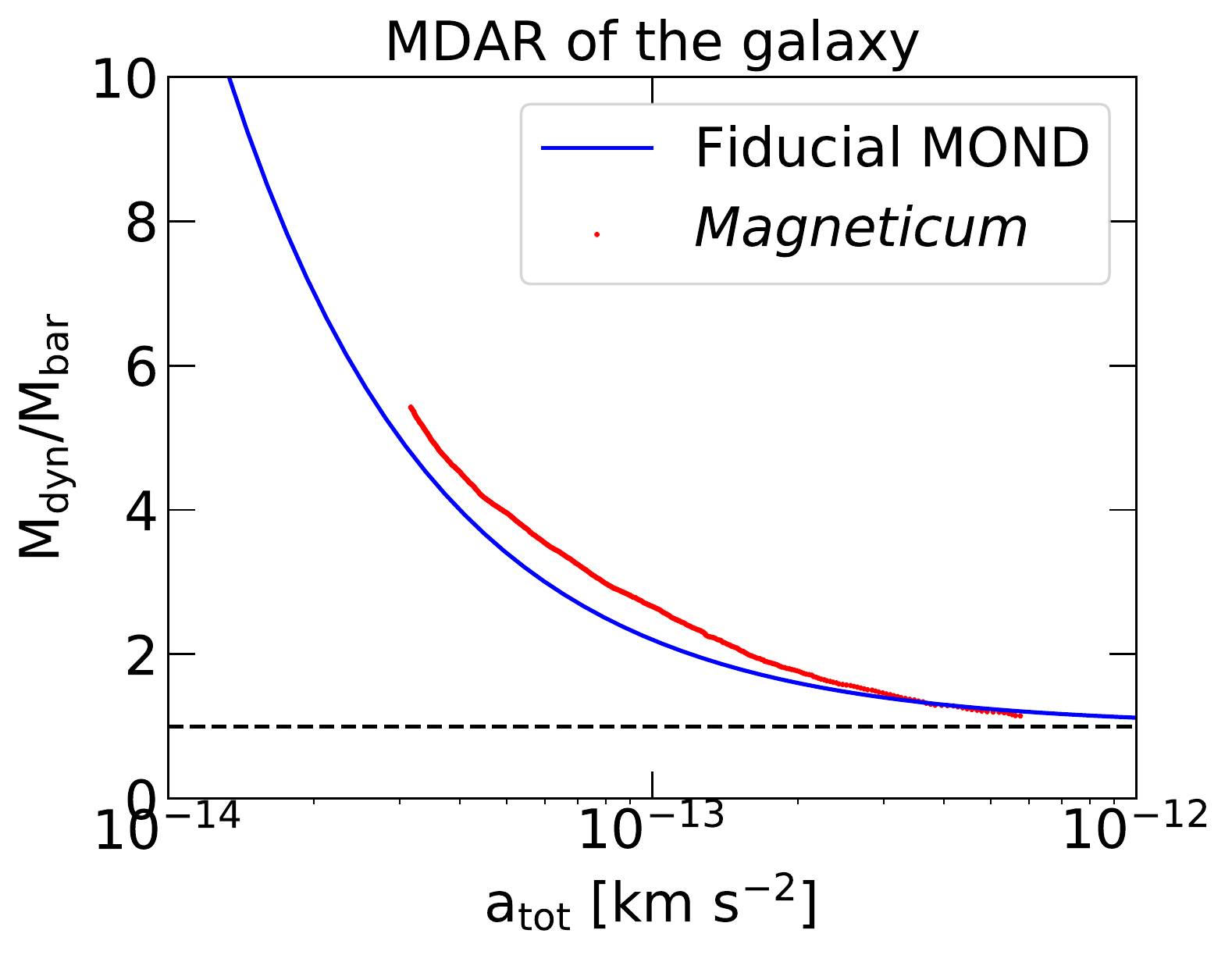}
\includegraphics[scale=0.25]{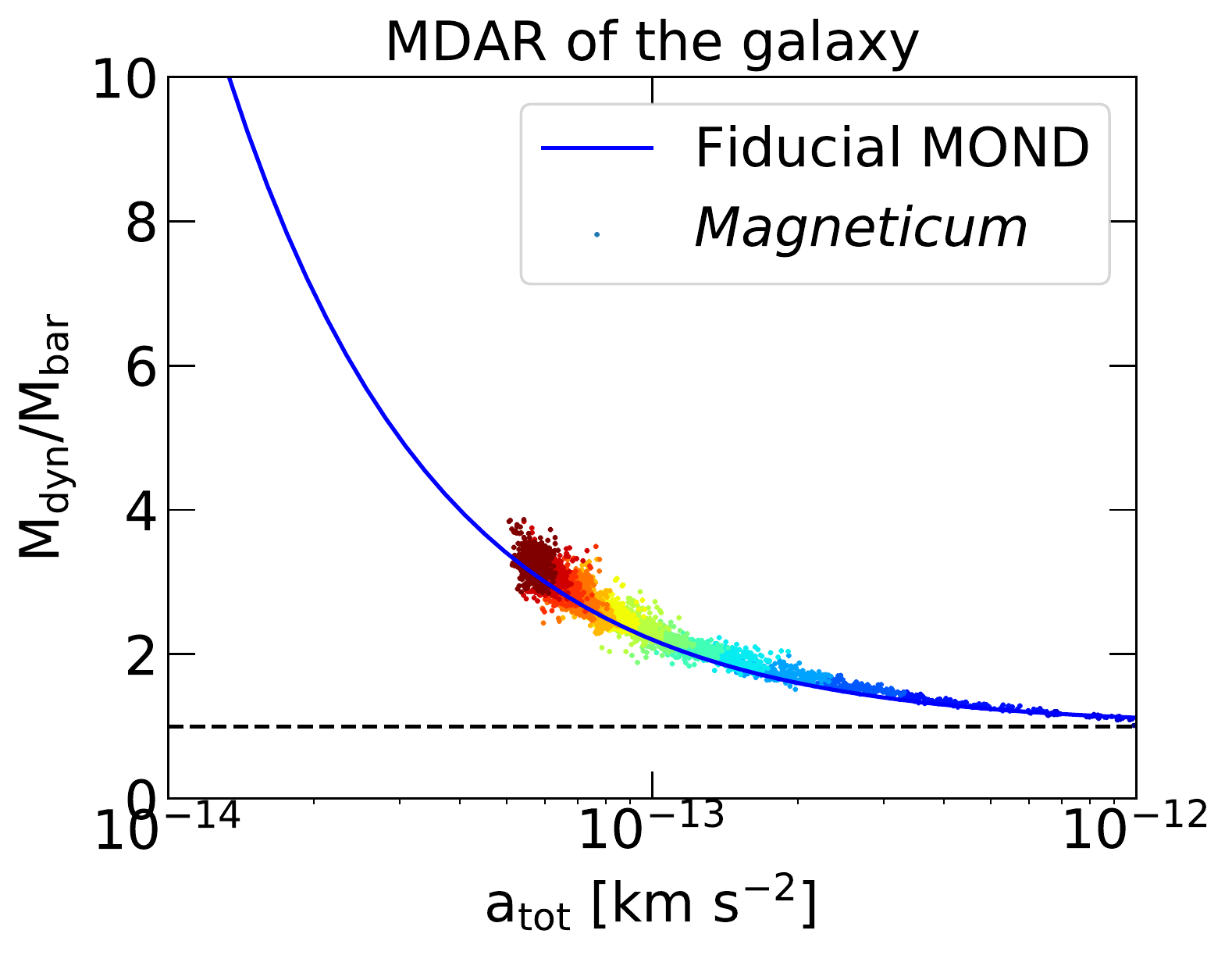}
\includegraphics[scale=0.25]{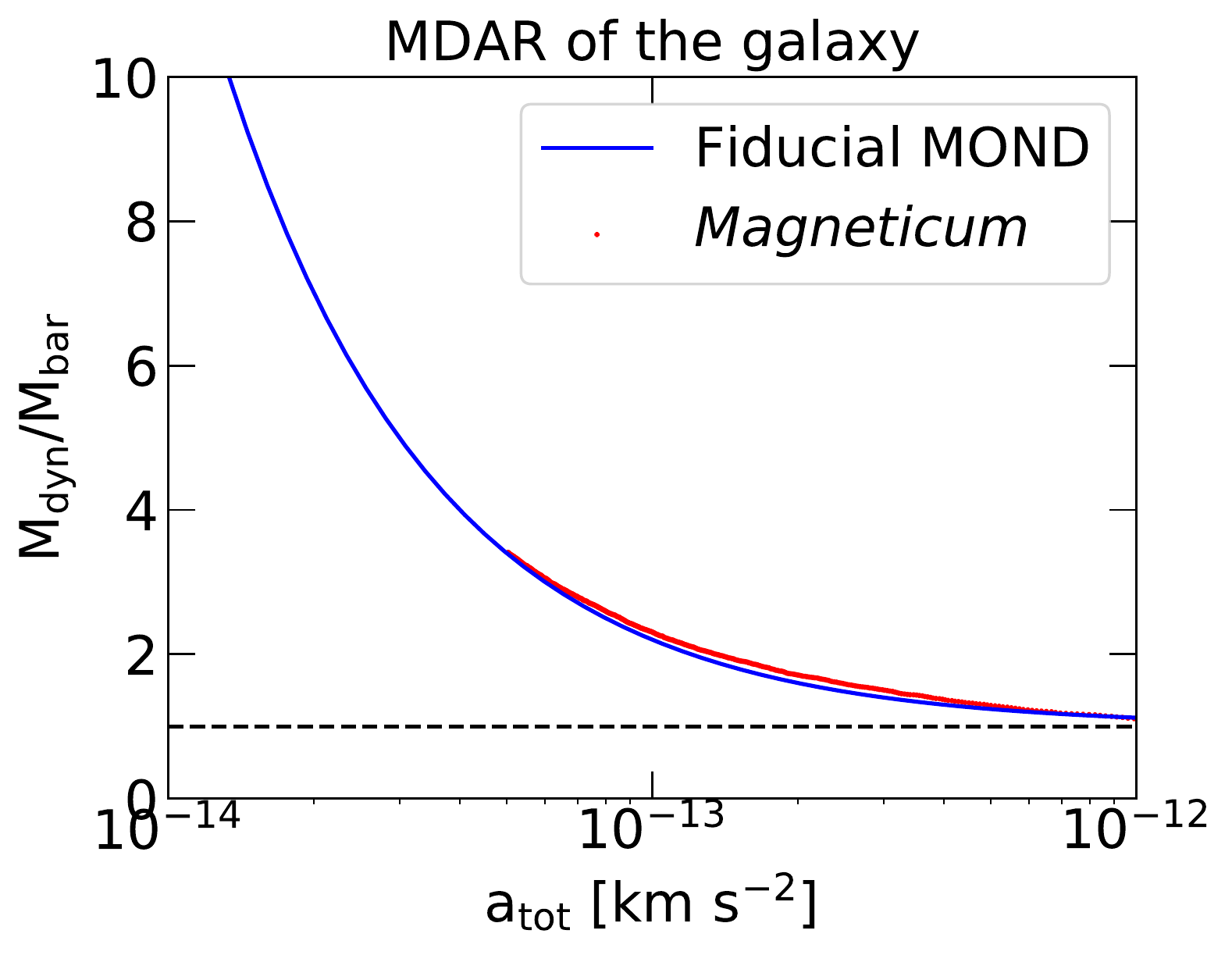}
\includegraphics[scale=0.28]{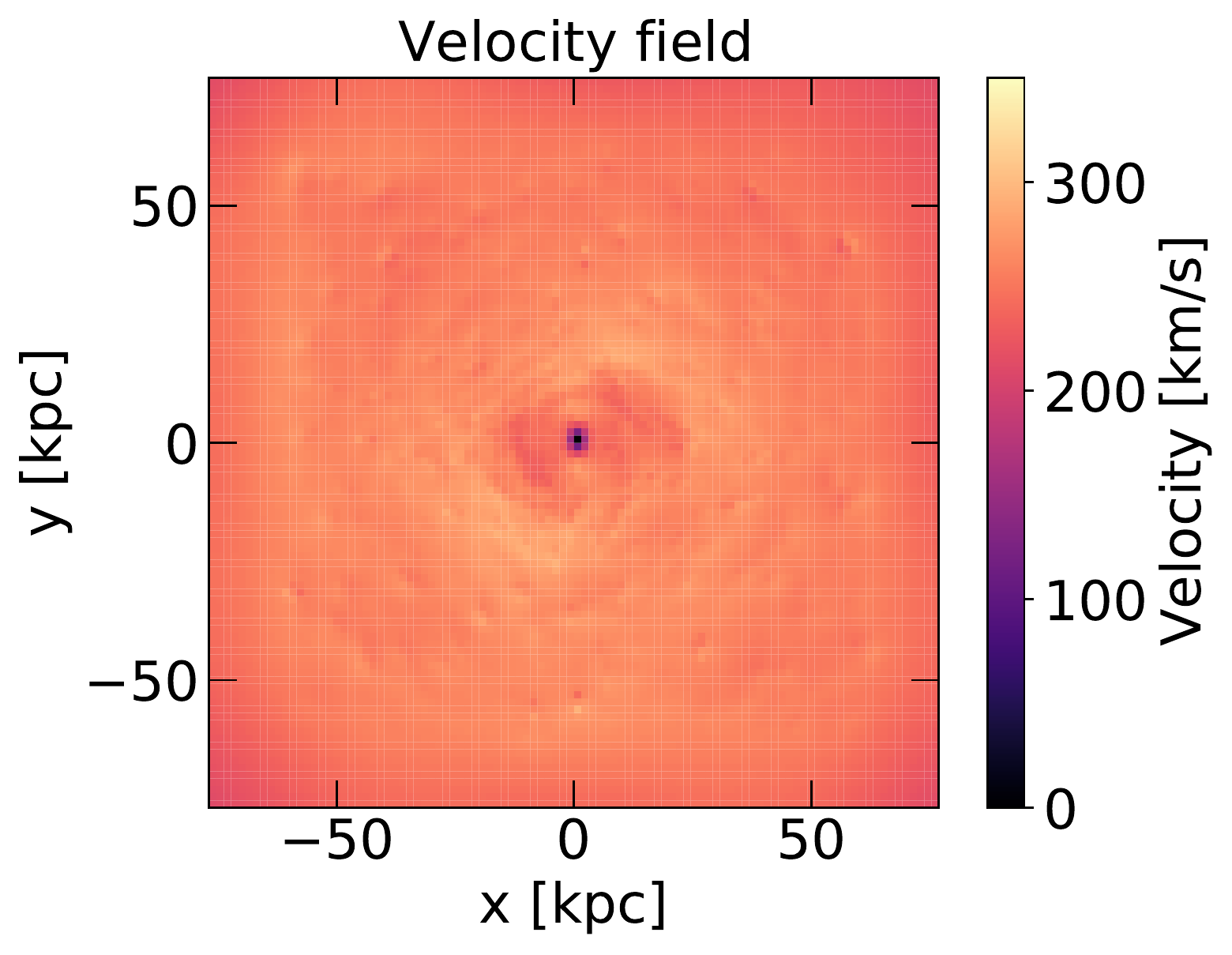}
\includegraphics[scale=0.28]{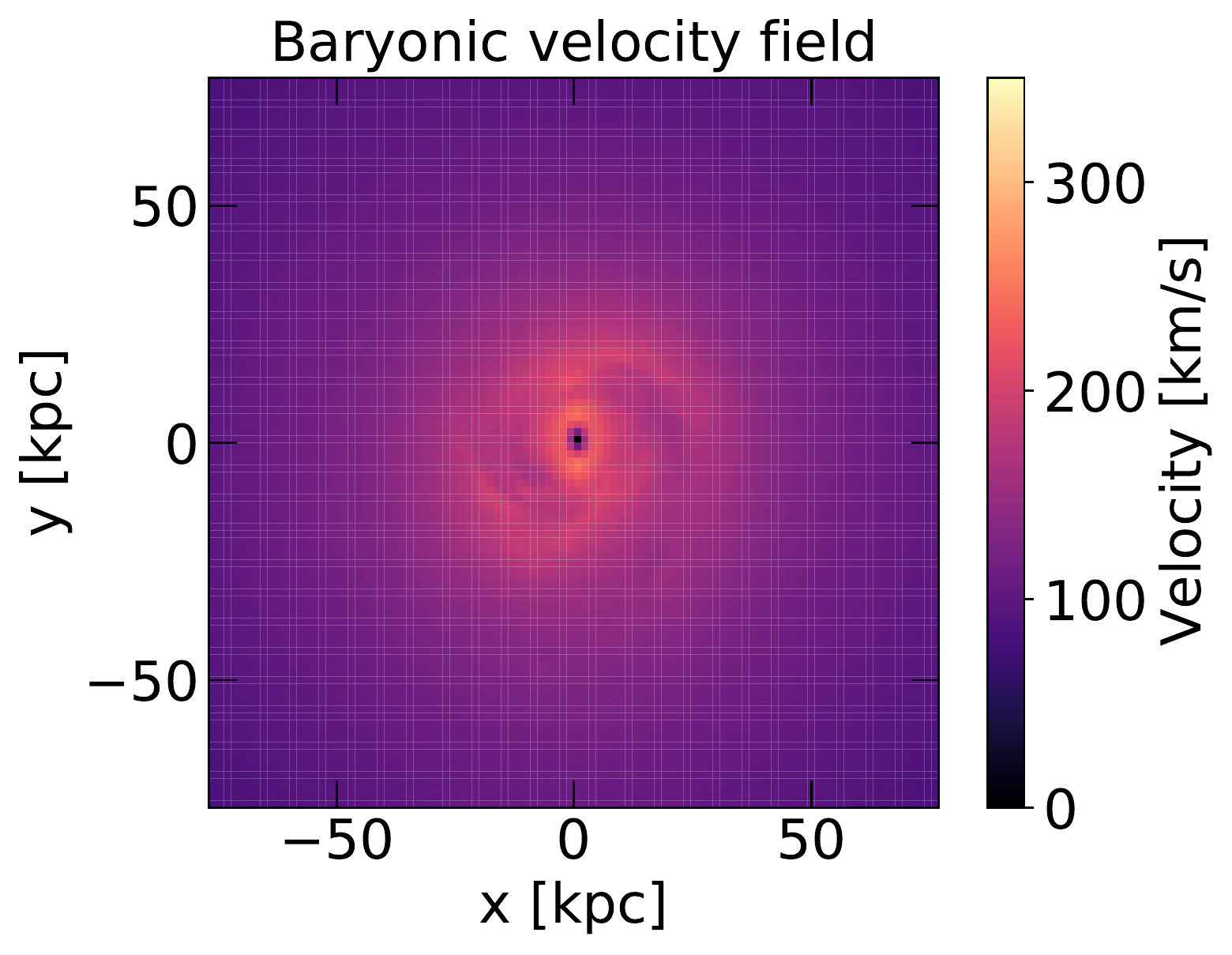}
\includegraphics[scale=0.28]{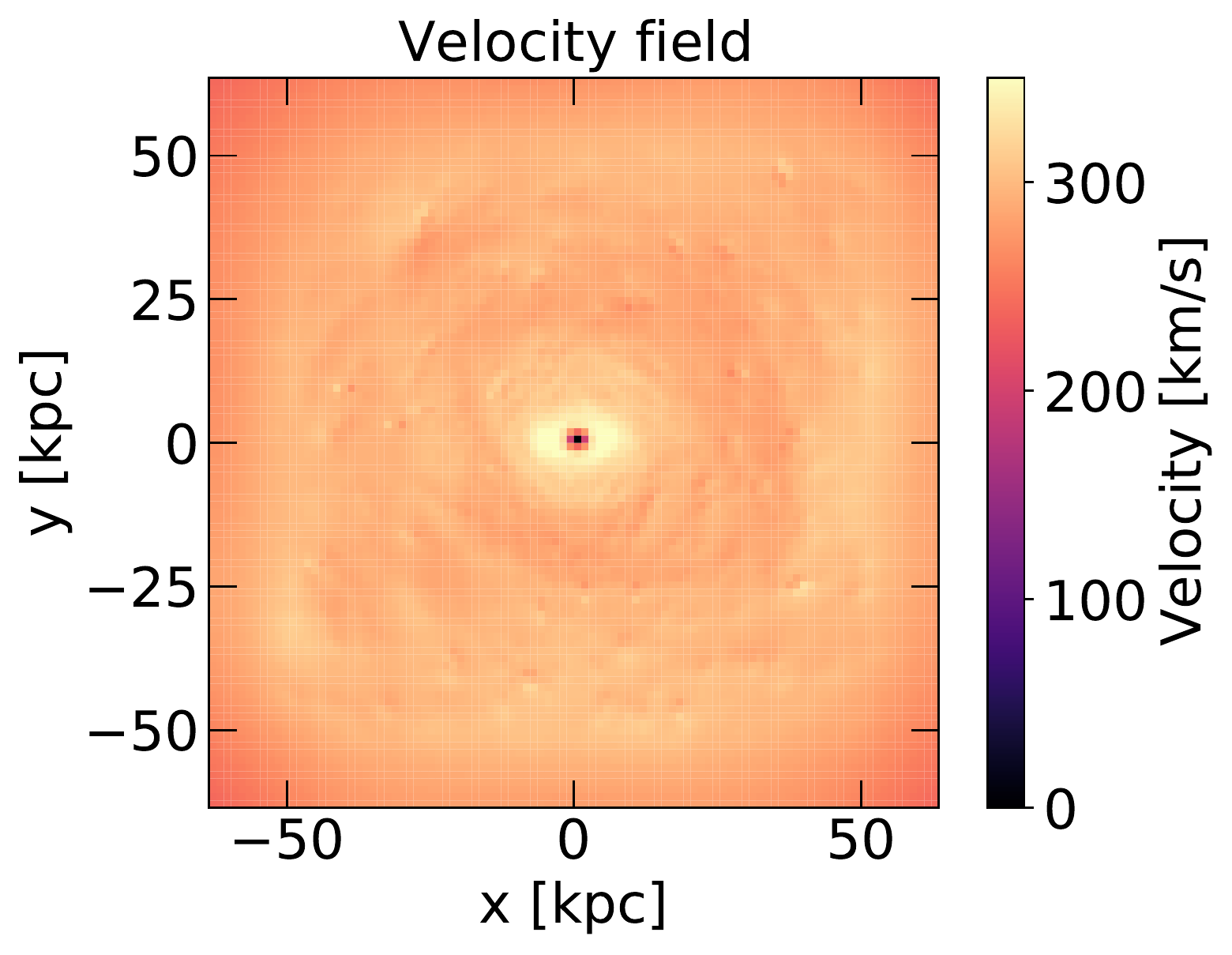}
\includegraphics[scale=0.28]{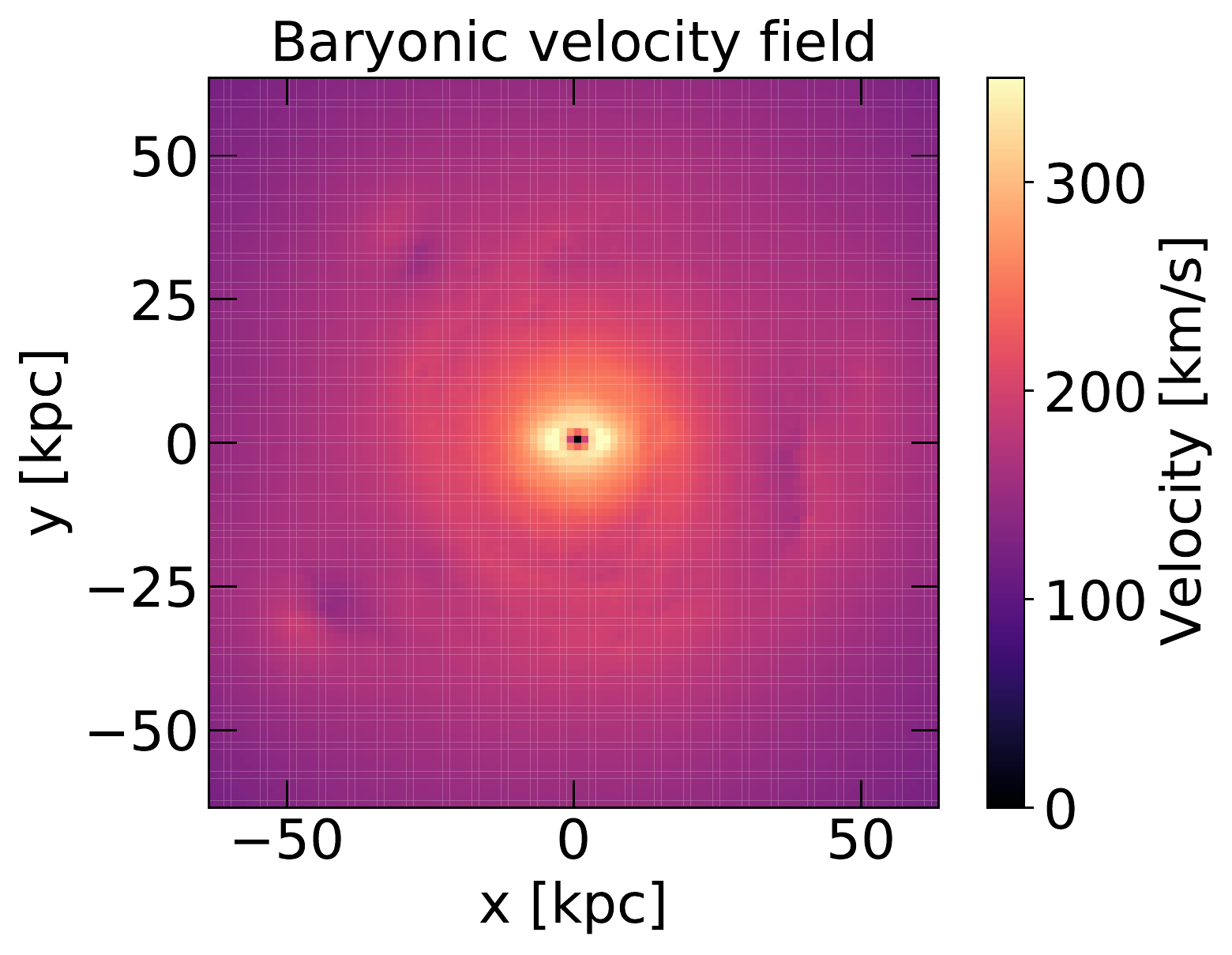}
\caption{The 4 top panels show a comparison of the MDAR from potential theory (left) and spherical approximation (right) for two \textit{Magneticum} galaxies. Colours on the left hand side indicate the different radial bins (increasing radius from blue to red), chosen as for the spherical approximation before.  Below that are the corresponding total and baryonic velocity maps.}
\label{compare}
\end{figure}

To get an idea for the intrinsic variation of the MDAR/RAR in a DM universe and in order to test the impact of using the earlier approximation, we will now depart from formula \eqref{GMoverR} that assumes the galaxy to be perfectly spherical. As this cannot possibly be the case for a disk-shaped galaxy, this method of obtaining values for the acceleration due to the mass distribution is not strictly accurate. To test for possible bias and noise that result solely from the use of this approximation, we used a relaxation scheme to obtain values for the gravitational potential $\Phi$ from the mass distribution of a galaxy, from which we proceeded to calculate the radial acceleration via
\begin{equation}
a(r)=-\frac{\partial\Phi}{\partial r} \label{potential}
\end{equation}
In Figure \ref{compare}, a comparison between the MDAR from the simplification from equation \eqref{GMoverR} (right panels) and the more accurate calculation based on formula \eqref{potential} (left panels) is shown for two galaxies. The intrinsic scatter predicted by the potential-based approach can be seen clearly. The presence of such scatter is not surprising, as this method captures non-axisymmetric substructure of the galaxy (such as spiral arms), while the spherical approximation does not. The important point for the extraction of the MDAR and RAR is, however, the similarity in the general shape of the profiles from the two methods: The data points of the MDAR from the potential scatter around the approximate MDAR. Therefore, formula \eqref{GMoverR} provides a sufficiently good approximation for the MDAR and RAR of \textit{Magneticum}, and likely does so in general.

\begin{figure}
\includegraphics[scale=0.5]{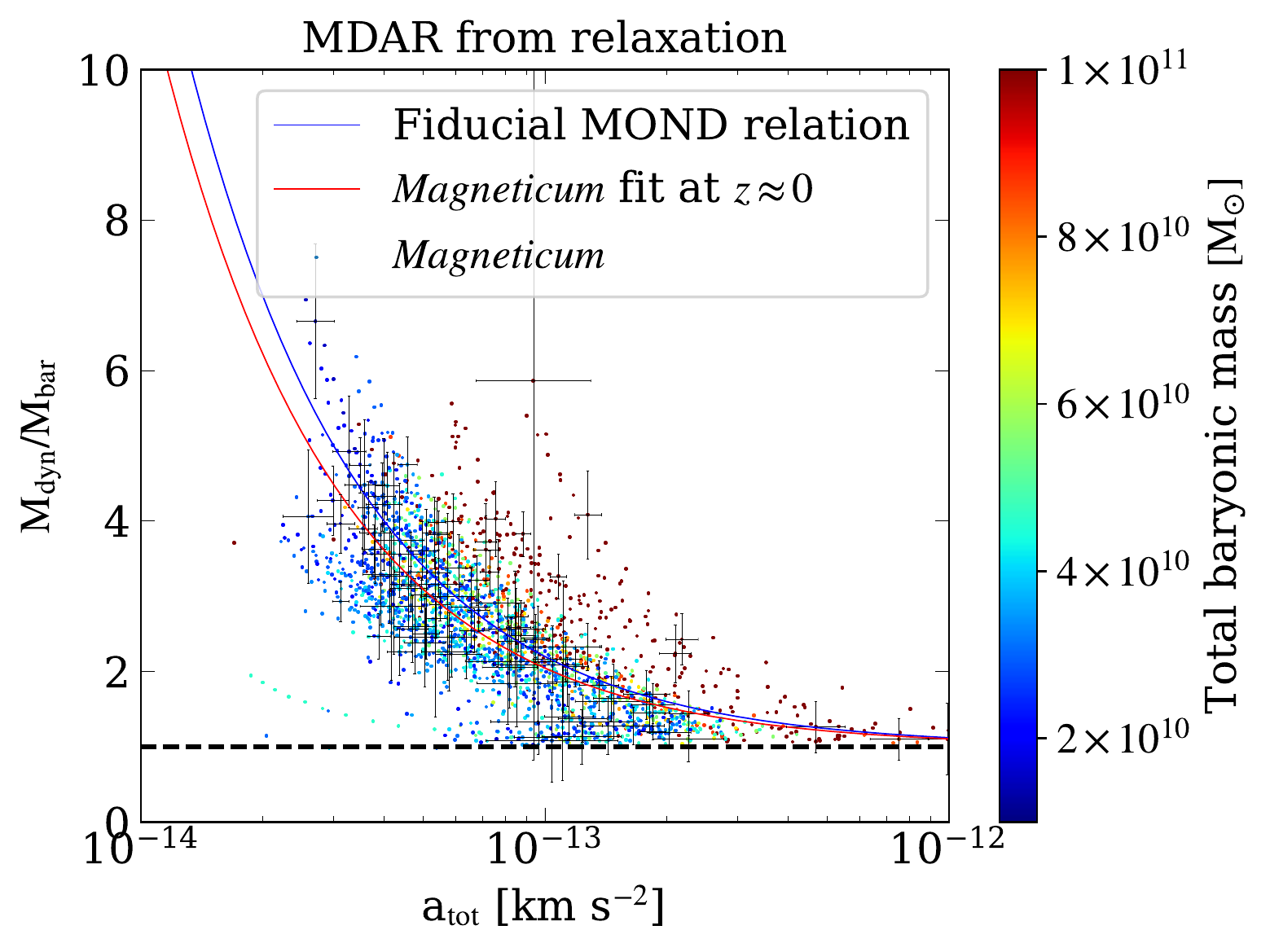}
\includegraphics[scale=0.5]{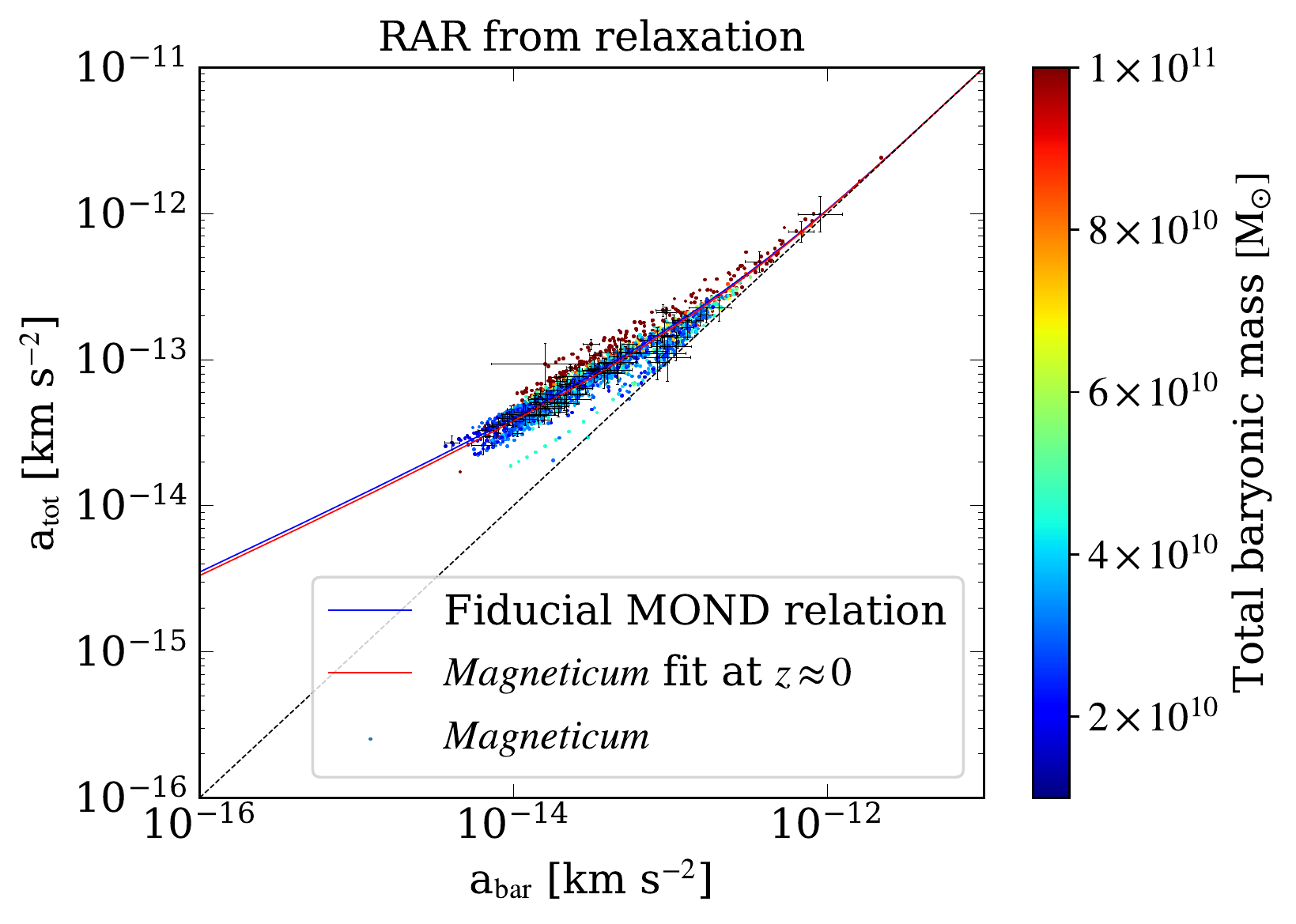}
\caption{RAR/MDAR from \textit{Magneticum} using potential theory with standard deviations indicated by errorbars, which are attached to 100 randomly selected points and correspond to the same points on the upper and lower panel. 11 of the total 2265 points were excluded due to their high standard deviation (larger than the average value).}
\label{poissonrarmdar}
\end{figure}

Figure \ref{poissonrarmdar} shows the result of applying this method to all disk galaxies at $z=0.1$. The errors were calculated as the standard deviation of the points within the respective radial bin. One can clearly see that this is very similar to the lower right panel in Figure \ref{MDAR_z}, validating the use of the spherical approximation in this statistical view as well. We note that the trend of galaxies with higher mass showing higher mass discrepancies is also still present here. The resulting values for $a_0$ are slightly smaller than before at this redshift.

\subsection{Increased sample size}
There is also the possibility of the galaxy sample still being too small to obtain a statistically significant sample, or even of a bias arising from limited box size. For this reason, in addition to the main sample from \textit{Box4/uhr}, we also used a bigger sample for redshift $z \approx 2$ from \textit{Box3/uhr}, which has a cosmological volume of (128 Mpc$h^{-1}$)$^3$ at the same resolution level evolved with a slightly updated black hole treatment \protect\citep[for details, see][]{Steinborn15}, reaching a redshift of $z = 2$.
Using this much larger sample we confirm that the relative difference of the best-fitting value for $a_{0}$ in the smaller box from the one in the larger box is no bigger than $\simeq$6 \% for both the RAR and the MDAR (see also the detailed Figure \ref{biggerbox} in the appendix). The value in the larger box is clearly higher than the one at $z\approx 0$ (in the smaller box). Therefore, this difference of box sizes does seem to have little impact on the observation of a change of the characteristic acceleration scale over time in
\textit{Magneticum}.

\begin{figure*}
\includegraphics[scale=1]{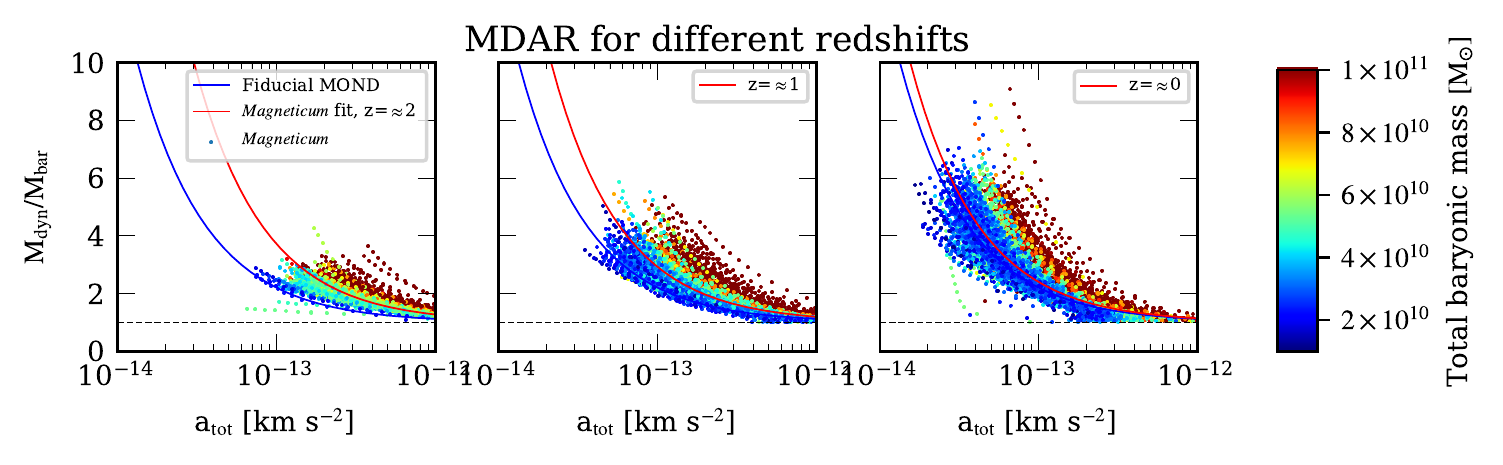}
\includegraphics[scale=1]{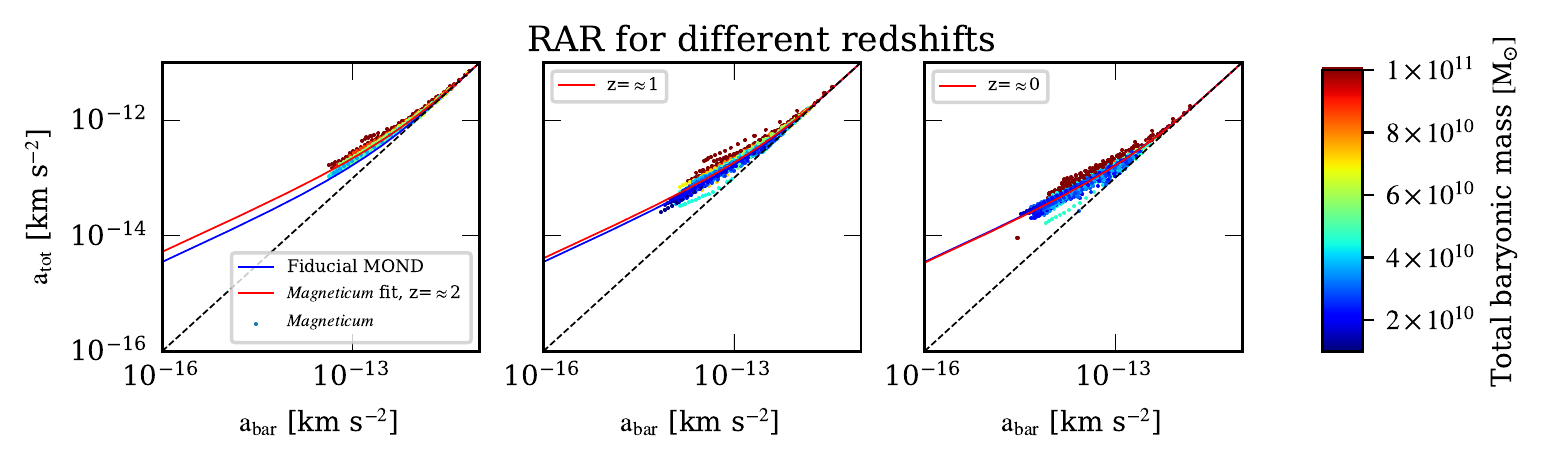}
\caption{MDAR and RAR of the spheroidal galaxies at $z \approx 0$,1,2. The similarity to the MOND prediction (blue lines) and therefore the MDAR and RAR can be clearly seen. The evolution with redshift is also similar to the one observed in \textit{Magneticum} disk galaxies, but $a_{0}$-values for spheroidal galaxies are larger. The mass trend in the MDAR and RAR (galaxies with higher stellar mass have higher mass-discrepancies) is even more pronounced here than in the disk galaxies.}
\label{spheroidals}
\end{figure*}

\subsection{MDAR/RAR in spheroidal galaxies}

In section \ref{Magneticum_simulation} it was discussed how the disk galaxies were selected from the total pool of galaxies at each redshift, using the $b$-value. This was done because typically the dynamical mass can be computed in a more straight forward way for rotation dominated systems than for elliptical galaxies, where the dynamics are dominated by random motions. However, in simulations like \textit{Magneticum} the acceleration data can always be calculated solely from mass distributions, which is of course possible for every kind of galaxy. Note that among others, \cite{2021MNRAS.506.5108C} demonstrated that the fraction of passive/quenched and star-forming galaxies within the \textit{Magneticum} simulations compares well to observations and the internal structures and dynamics also compare well to observations for the spheroidal galaxy population \citep{Remus16,Remus2017b,Schulze18,Schulze20}. As the existence of an MDAR/RAR-like relation reflects the different distribution of dark matter and baryons within the simulated galaxies, it is interesting to check if similar relations in principal also exist for different types of galaxies. Therefore, data for the spheroidal population of galaxies in \textit{Magneticum} at redshifts $z\approx 0,1,2$ is shown in Figure \ref{spheroidals}, calculated with the spherical approximation as for the disk galaxies before. Interestingly, the MDAR and RAR also clearly exist in these galaxies.

\begin{figure*}
\includegraphics[scale=1]{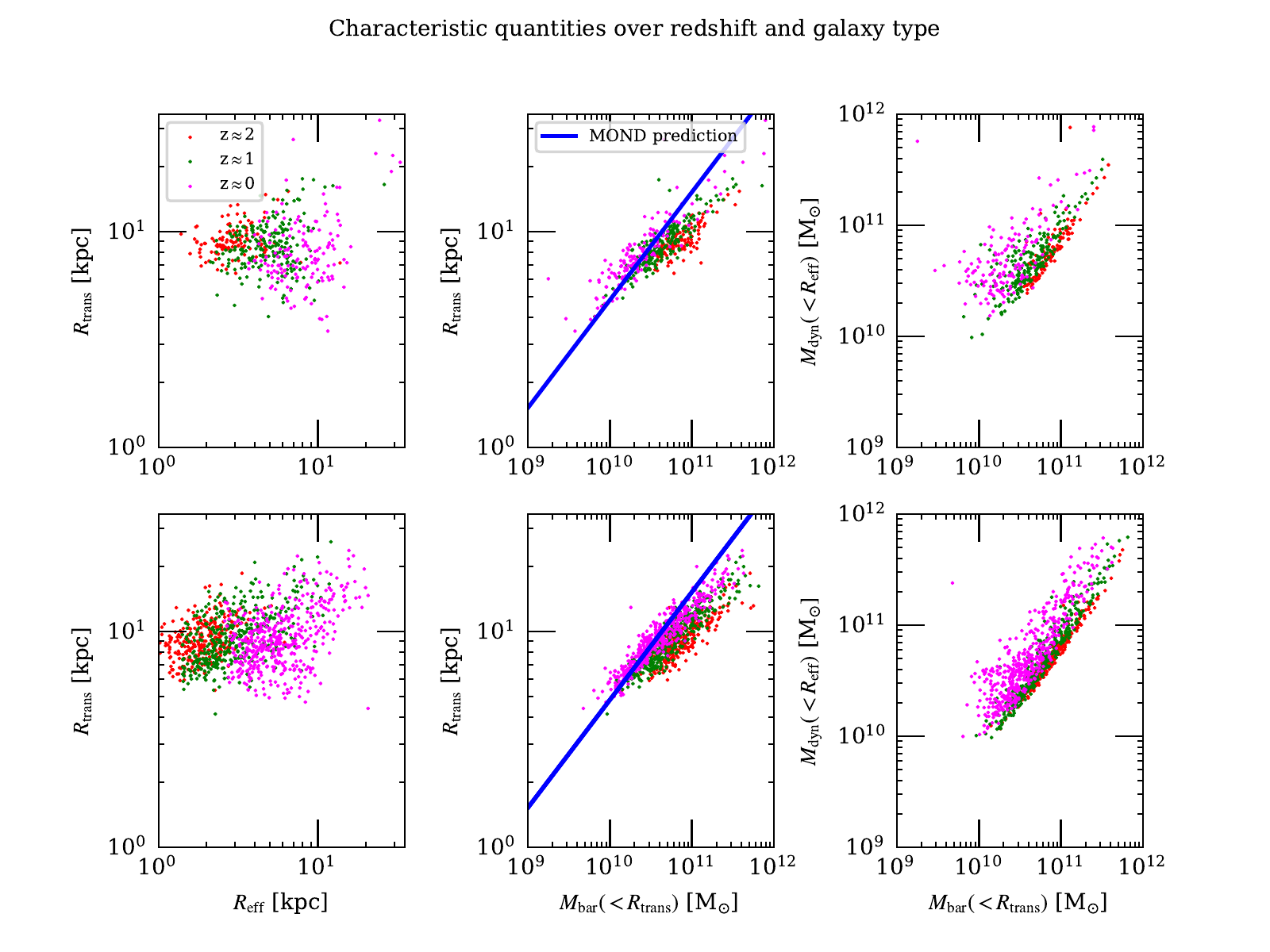}
\caption{Showcasing the correlation (and their evolution, color coded as indicated in the plot) of the \textit{effective radius} ($R_{\mathrm{eff}}$), the \textit{transition radius} ($R_{\mathrm{trans}}$), the baryonic mass inside $R_{\mathrm{trans}}$ ($M_{\mathrm{bar}}(<R_{\mathrm{trans}})$) and the total (dark+baryonic) mass inside $R_{\mathrm{eff}}$ ($M_{\mathrm{dyn}}(<R_{\mathrm{eff}})$) for the disk galaxies (top panels) and spheroidal galaxies (bottom panels).}
\label{quantities}
\end{figure*}

Although these predictions for spheroidal galaxies are generally harder to compare to observations than in the case of disk galaxies, the RAR in nearly spherical galaxies has been studied by \cite{2019ApJ...877...18C}. They obtain an empirical RAR of $\frac{a_{\mathrm{tot}}}{a_{\mathrm{bar}}}-1\approx\frac{a_0}{a_{\mathrm{bar}}}$ in the range $a_0 <a_{\mathrm{bar}} <100 \cdot a_0 $. This is similar in form to equation \eqref{MDARsimple}, but with baryonic instead of total acceleration on the right hand side, and might hint at such an underlying relation. On the other hand, other interpolation functions appear inconsistent with their data. Future studies could shed light on the regime $a_{\mathrm{bar}}<a_0$ as well to provide further data for comparison. \cite{2020ApJ...903L..31C} also conclude the general existence of a universal acceleration scale in elliptical galaxies. The results for spheroidal galaxies are illuminating, as they show that the underlying physics leading to the MDAR/RAR in a $\Lambda$CDM universe are not exclusive to disk galaxies but reflect more how the different matter components in principle are distributed within simulated galaxies in the $\Lambda$CDM framework. \cite{2019arXiv191006345G} discuss the importance of stellar feedback processes on the universal acceleration scale $a_{0}$ and come to the conclusion that the existence of such a characteristic scale in a $\Lambda$CDM universe is a result of the dependence of feedback processes on the surface density in disk galaxies, because this density also sets the acceleration scale in any particular galaxy (or region of a galaxy). Specifically, $a_{0}$ is argued to be a threshold above which star formation becomes efficient. At higher densities/accelerations, baryonic matter accumulates, which results in low mass discrepancies. Considering the existence of an MDAR/RAR in spheroids in \textit{Magneticum}, however, star formation alone does not seem to be a sufficient explanation. This is because very little star formation takes place in elliptical galaxies in nature, which is also reflected in the spheroidal galaxies within the \textit{Magneticum} simulations. 

Instead, it is thought that a significant fraction of ellipticals are a result of major mergers of (spiral) galaxies \citep[see e.g.][]{1977egsp.conf..401T,1978MNRAS.184..185W,1996ApJ...471..115B} or series of multiple minor mergers \citep[see e.g.][]{2003ApJ...590..619M,2009ApJ...699L.178N,2012ApJ...754..115J} with only long-lived stars remaining in the galaxy after enough time has passed. 
It does seem unlikely that the MDAR/RAR would somehow naturally get transferred from colliding (spiral) galaxies to the elliptical they merge into, as the formation history is very different from elliptical to elliptical, where the different formation channels depend on the merger configurations such as orbital parameters, the merger sequence (e.g. minor/major), intrinsic properties of progenitors (e.g. wet/dry) and involves drastically varying numbers and sizes of progenitor galaxies. These different formation pathways were supported by the findings in observations of elliptical galaxies exhibiting different rotation patterns \citep[see e.g.][]{Emsellem2007}, which was also shown for the spheroidal galaxies in the \textit{Magneticum} simulations \citep[see e.g.][]{Schulze18}. 

Considering this, the MDAR/RAR in spheroidal galaxies is likely to be either a consequence of the feedback processes that are still very active in elliptical galaxies (such as AGN and supernovae), the simple cooling of gas, or is the result of the fundamental behaviour of spheroidal galaxies, where their complex dynamical interaction between the baryonic and dark matter component leads to a co-evolution of total density profiles and the dark matter fraction \citep{Remus2017b}. \\

\subsection{Characteristic quantities of the mass distribution}
Finally, we take a closer look at the mass distribution of the different components within the simulated galaxies from the \textit{Magneticum} simulations in the context of MOND. Specifically, we will investigate a possible correlation between parameters of those galaxies that characterize the distribution of baryonic and/or dark matter. One of these is the \textit{effective radius} $R_{\mathrm{eff}}$, inside which half of the total baryonic mass of the galaxy is located. As a characteristic radius for the dark matter, we instead choose the \textit{transition radius} $R_{\mathrm{trans}}$, which is the smallest radius such that the enclosed dark mass is larger than the baryonic mass. Further, we considered the baryonic mass inside $R_{\mathrm{trans}}$, denoted $M_{\mathrm{bar}}(<R_{\mathrm{trans}})$ and the total (dark+baryonic) mass inside $R_{\mathrm{eff}}$, denoted $M_{\mathrm{dyn}}(<R_{\mathrm{eff}})$. Figure \ref{quantities} displays the relation between these quantities for the redshifts $z\approx0,1,2$ for both disk and spheroidal galaxies in \textit{Magneticum}. 

It can be clearly seen in the left panels that the two characteristic radii do not show a strong correlation in \textit{Magneticum} galaxies. A correlation between $R_{\mathrm{trans}}$ and $M_{\mathrm{bar}}(<R_{\mathrm{trans}})$, as shown in the middle panels, is predicted by MOND through the MDAR by definition of the transition radius
\begin{equation}
M_{\mathrm{bar}}(<R_{\mathrm{trans}})=\frac{a_{0}}{2 \cdot G} \cdot R_{\mathrm{trans}}^{2}.
\end{equation}
This MOND prediction is also displayed in the figure, assuming no redshift evolution in $a_{0}$. The slope is slightly steeper in \textit{Magneticum} than predicted by MOND, but a correlation appears to exist.

The right panels of Figure \ref{quantities} show a close relationship between $M_{\mathrm{bar}}(<R_{\mathrm{trans}})$ and $M_{\mathrm{dyn}}(<R_{\mathrm{eff}})$. Indeed, the slopes are close to 1, indicating that these masses are actually almost equal (on average) over the redshift range from 2 to 0 and between different galaxy types. Since $R_{\mathrm{eff}}$ is determined by the distribution of baryonic matter and $R_{\mathrm{trans}}$ by that of all forms of matter, this 'cross-relationship' (baryonic mass inside radius characterized by total mass distribution and total mass inside radius characterized by baryonic mass distribution are close to equal) could be a fundamental manifestation of the interplay between baryonic and dark matter and therefore of the MOND-phenomenology in $\Lambda$CDM. Interestingly, the spheroidal galaxies show tighter relations than the disk galaxies, however larger sample size might be needed for a solid, quantitative measurement of the scatter. Note that generally the baryonic and dark matter content of the spheroidal galaxies from the \textit{Magneticum} simulations have shown to compare well to observations out to large distances \citep{2020ApJ...905...28H}.

\subsection{Comparison with MOND prediction for $a_{0}(z)$}
\label{mond_pred}
When trying to obtain a prediction for the evolution of $a_{0}$ with $z$ from MOND, it is important to keep in mind that MOND is not a relativistic theory. It is therefore unlikely to provide an appropriate description of cosmological phenomena like a change in $a_{0}$ over cosmic time. A relativistic theory with MOND as a limiting case, such as Tensor Vector Scalar Gravity (TeVeS) (\cite{2004PhRvD..70h3509B}), should preferably be used. Nonetheless, it has been pointed out since the inception of MOND (\cite{1983ApJ...270..365M}) that $a_{0}\approx cH_{0}$, hinting at the possibility that $a_{0}$ should have a similar redshift dependence as $H$. \cite{2008PhRvD..77j3512B} argue that by setting $\Omega_{m}=0.25$, $\Omega_{\Lambda}=0.75$ and assuming zero curvature, the matter-domination throughout the relevant period leads to the prediction
\begin{equation}
a_{0}(z)\approx a_{0}(0)\cdot(1+z)^{\frac{3}{2}} \label{MONDverynaive}
\end{equation}
for $z>1$. However, as they point out, this approach is not complete - $\Omega_m$ should really only include the baryon fraction. Including the effect of $\Omega_\Lambda$, one obtains 
\begin{equation}
a_{0}(z)\approx a_{0}(0)\cdot [\Omega_m (1+z)^3 +\Omega_\Lambda ]^{\frac{1}{2}} \label{MONDnaive} \, .
\end{equation}
There does not seem to exist a consensus for the prediction of $a_0$ from MOND, and the above formula should only be seen as one possibility. In Figure \ref{a0(z)}, one can see that this relation fails to accurately describe the trend observed in \textit{Magneticum}, with the change in \textit{Magneticum} being somewhat slower as redshift increases. The strong redshift-scaling in equation \eqref{MONDverynaive} seems to be inconsistent with observational data (\cite{2017arXiv170306110M}).

Based on a subset of TeVeS models, it is argued instead that changes in $a_{0}$ should happen on time scales \textit{much larger than Hubble's}, although an observable change on cosmological timescales is not excluded by TeVeS (\cite{2008PhRvD..77j3512B}). In observations, this prediction would be clearly distinguishable from the $a_{0}(z)$-relation form this work with $a_{0}(z=2)\approx3\cdot a_{0}(z=0)$. Another relativistic completion of MOND is \textit{Covariant Emergent Gravity}, in which the redshift-dependence of $a_{0}$ is related to the size of the cosmological horizon (\cite{2018IJMPD..2747010H}). It is pointed out that this approach gives rise to an increase of $a_{0}$ with greater $z$, but that this dependence is much smaller than the one found in simulated galaxies from the MUGS2-sample by \cite{2017ApJ...835L..17K}.

\subsection{Comparison to other simulations and models}

The existence of an MDAR and RAR obtained from a large galaxy sample from the \textit{Magneticum} simulations echoes earlier findings in other, individual galaxy simulations, such as those described by \cite{2019MNRAS.485.1886D}, where the set of NIHAO galaxy formation simulations was used. Further, \cite{2017PhRvL.118p1103L} demonstrate the existence of acceleration-scaling similar to that predicted by MOND in EAGLE. \cite{2016MNRAS.456L.127D} used a semi-empirical model to construct galaxies with appropriate halos and it is found that both the MDAR and BTFR can be obtained in this way, although the latter still shows an exponent of less than 4. Using this model, galaxies with higher total stellar mass tend to have larger mass discrepancies than those with lower mass (Figure 3 of their Letter), similar to the results in this work. \\
\cite{2021MNRAS.507..632P} show how the existence of acceleration relations in a $\Lambda$CDM universe follows via analytical formulas for the profiles of DM halos resulting from adiabatic contraction under the influence of baryonic feedback processes. These arguments are tested and confirmed using a ``baryonification" algorithm for a mock-catalogue described in more detail in \cite{2021MNRAS.503.4147P}, which is however limited to low redshift. These results indicate that the existence of the RAR/MDAR in both disk and spheroidal is a natural phenomenon in a universe with CDM. \\
\cite{2017MNRAS.464.4160D} and  \cite{2017MNRAS.472L..35D} investigate the scatter in the MDAR and BTFR via halo abundance matching from mock galaxies, concluding that the scatter predicted from $\Lambda$CDM exceeds the observed one for both, even though the shape of the MDAR is indeed reproduced. \\ 
The existence of a decrease of the parameter $a_{0}$ with declining redshift also agrees with the trend observed by \cite{2017ApJ...835L..17K} using the MUGS2 simulations.

\section{Conclusions}

We used galaxies from the \textit{Magneticum} simulation to probe the existence of acceleration relations in $\Lambda$CDM similar to those predicted by Modified Newtonian Dynamics, the MDAR and RAR as well as the existence of a BTFR. The BTFR in \textit{Magneticum} does show an exponent close to 3 over a wide range of redshifts, in contrast to exactly 4 predicted by MOND. It is also found that the set of \textit{Magneticum} galaxies are showing relations similar to the MOND MDAR and RAR at redshifts $z\approx0-2.8$. However, the characteristic acceleration scale $a_{0}$ is seen to decrease over cosmic time, with a decrease of more than a factor of 3 from $z \approx 2.8$ to $z \approx 0$. Both the existence of the MDAR/RAR as well as this redshift-trend are also observed in elliptical galaxies. 

As argued by \cite{2015CaJPh..93..250M}, the observation of acceleration scaling relations in the form of the MDAR and RAR did not long ago seem like a promising test to distinguish between $\Lambda$CDM and MOND. The reason is that this observation was predicted by MOND more than 30 years ago, before sufficiently accurate data became available, while there was no obvious reason why these relations should exist in a $\Lambda$CDM universe. However, considering the results of this work as well as those of earlier work referenced herein, it does seem like acceleration relations do arise naturally in $\Lambda$CDM simulations, and therefore in a $\Lambda$CDM universe. Thus, the mere existence of the MDAR and RAR in the Universe is not sufficient to distinguish between $\Lambda$CDM and MOND. Accordingly, the focus should perhaps be shifted to the specifics of the MDAR and RAR, like the magnitude of scatter in different acceleration ranges, as is done in detail by \cite{2019MNRAS.485.1886D}. From the evolution $a_{0}(z)$ found in this work, as well as the work by \cite{2017ApJ...835L..17K}, the redshift-dependence of the universal acceleration scale also seems to be a promising candidate to distinguish between $\Lambda$CDM and MOND observationally, which has been argued before (\cite{2018IJMPD..2747010H}). 

Especially here, the large variation of $a_0$ between $z\approx 0$ and $z \approx 3$ as predicted by the galaxies from the \textit{Magneticum} simulations offers an important opportunity for the next generation of observations to shed new light on the fundamental differences between MOND and the standard CDM paradigm.

\section*{Acknowledgements}

This work was supported by the Deutsche Forschungsgemeinschaft (DFG, German Research Foundation) under Germany’s Excellence Strategy - EXC-2094 - 390783311. KD acknowledges support by the COMPLEX project from the European Research Council (ERC) under the European Union’s Horizon 2020 research and innovation program grant agreement ERC-2019-AdG 882679. The calculations for the hydrodynamical simulations were carried out at the Leibniz Supercomputer Center (LRZ) under the project pr83li. We are especially grateful for the support by M. Petkova through the Computational 
Center for Particle and Astrophysics (C2PAP).

\section*{Data Availability}
Public data releases exist for \textit{Magneticum} \citep{Ragagnin2017}, and can be found at {\tt http://magneticum.org/data.html} respectively. The data presented in this work are not part of the public catalogs but can be provided on reasonable request. 




\bibliographystyle{mnras}
\bibliography{MONDbib}



\appendix

\section{RAR plots}

\begin{figure}
\includegraphics[scale=0.6]{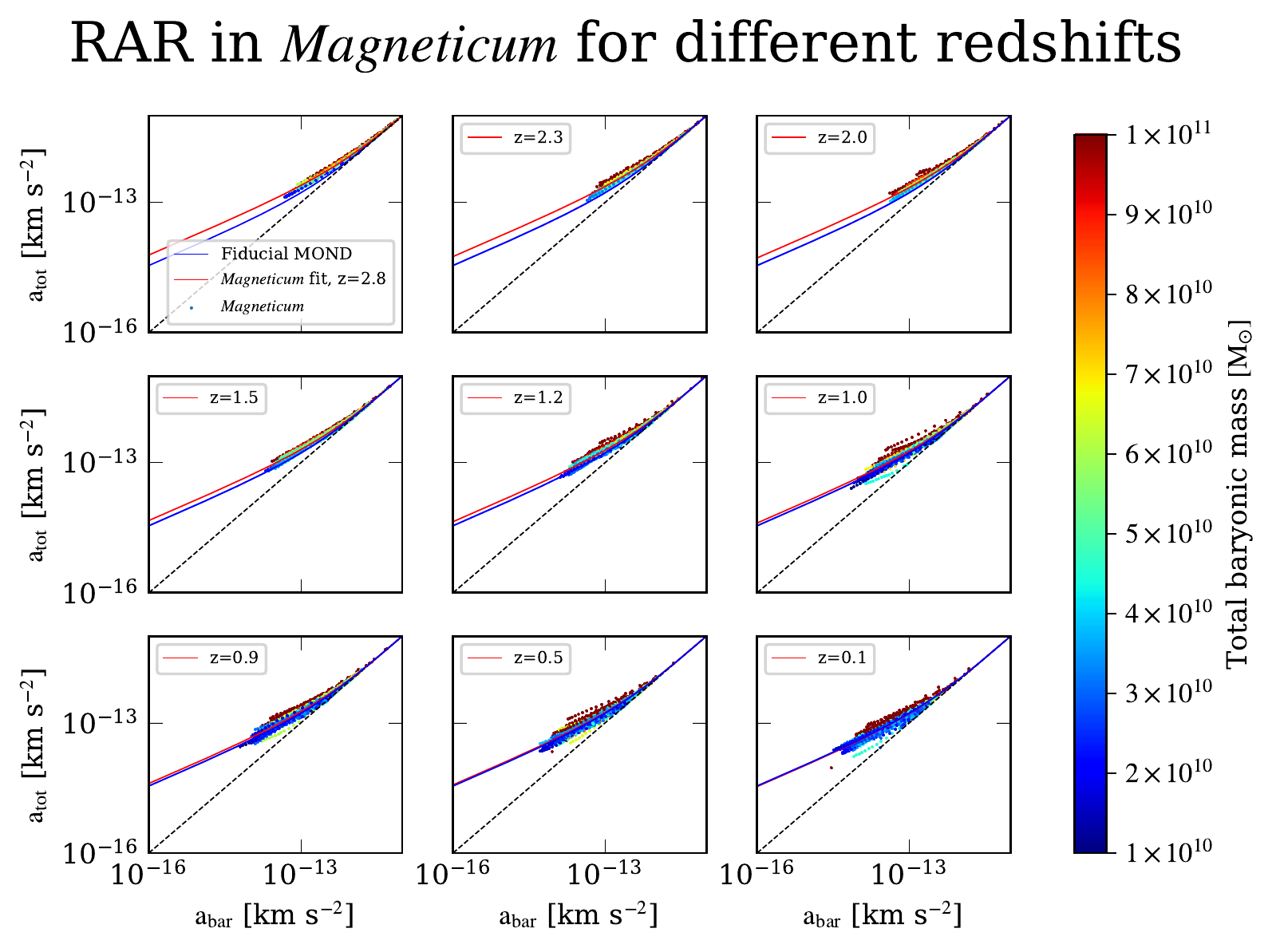}
\caption{RAR in \textit{Magneticum} at different redshifts. Points are from \textit{Magneticum} galaxies from each particular redshift and their colour indicates the total stellar mass of the galaxy from which a given data point is taken. The function represented by the blue curve is from equation \ref{eq:a1} with a fiducial value of $a_{0}=1.2 \cdot 10^{-13}\mathrm{km} \, \mathrm{s}^{-2}$, while the function shown as the red curve has the same form, but uses $a_{0}$ as a fitting parameter.}
\label{RAR_z}
\end{figure}

As discussed earlier, the predictions of MOND can be equivalently formulated in form of the MDAR and RAR. The MDAR was chosen for analysis over different redshifts in the body of the work due to its better readability. To complete the comparison, the RAR is displayed in Figure \ref{RAR_z}.

\section{Composition of the galaxy sample in Magneticum}
\begin{figure}
\includegraphics[scale=0.5]{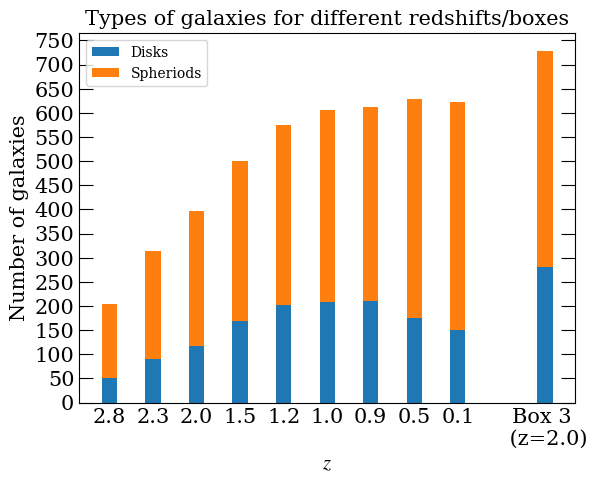}
\caption{Total number and composition of the galaxy sample at different redshifts, also including the numbers for \textit{Box3/uhr} at $z \approx 2.0$.}
\label{galaxyfigure}
\label{TFR}
\end{figure}

The sample of galaxies extracted from \textit{Box4/uhr} of the \textit{Magneticum} simulation typically contains around 150 disk and 400 spheroidal galaxies at the different redshifts used for the analysis, with significantly fewer galaxies in the two highest redshifts used. The additional sample extracted from the larger simulation box (e.g. \textit{Box3/uhr}) surpasses these numbers, even at high redshift. Figure \ref{galaxyfigure} displays both the relative and absolute number of disk and spheroidal galaxies used for the analysis at the different redshifts.

\section{Results from \textit{Box3/uhr}}

\begin{figure}
\includegraphics[scale=0.5]{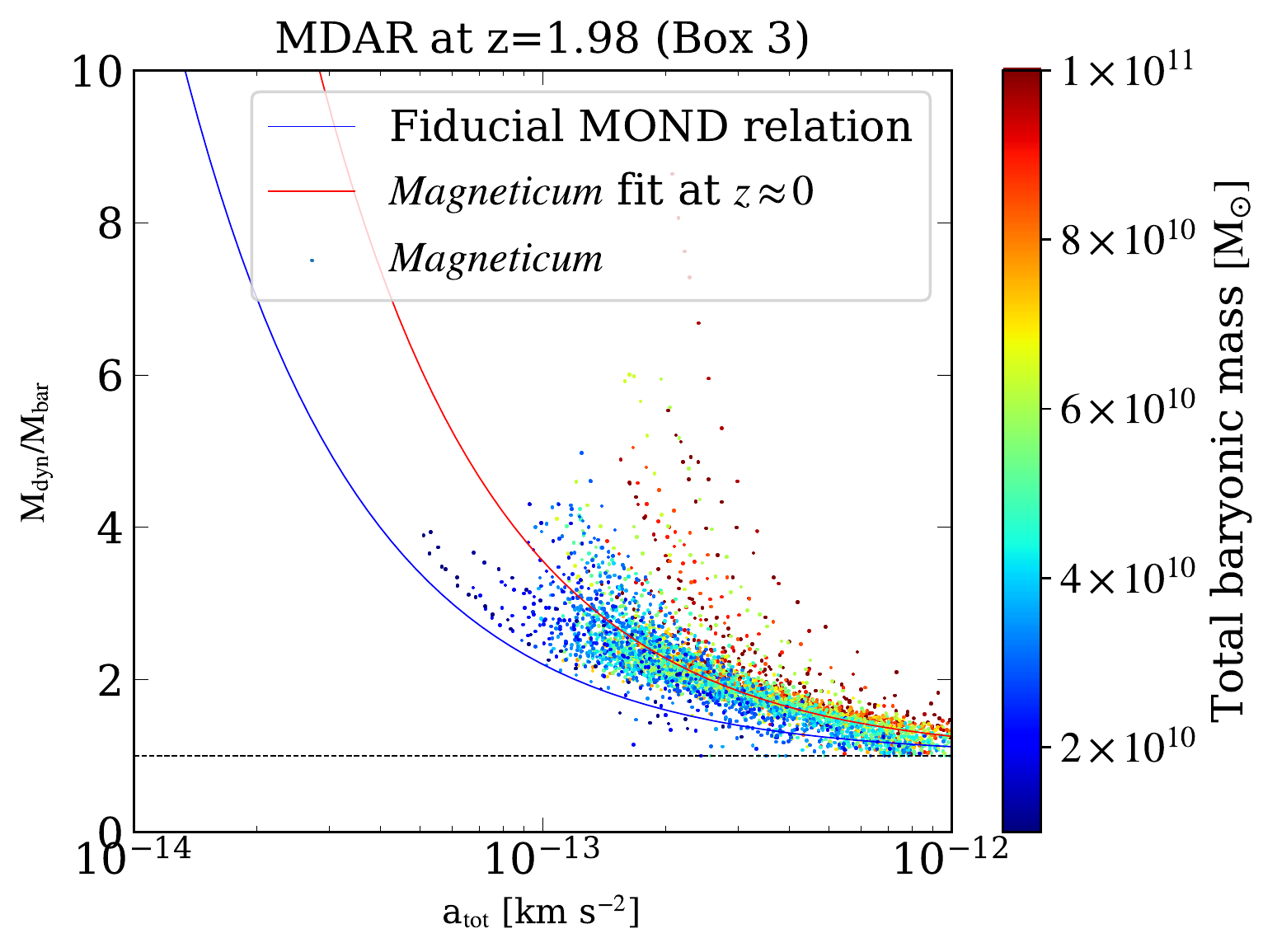}
\includegraphics[scale=0.5]{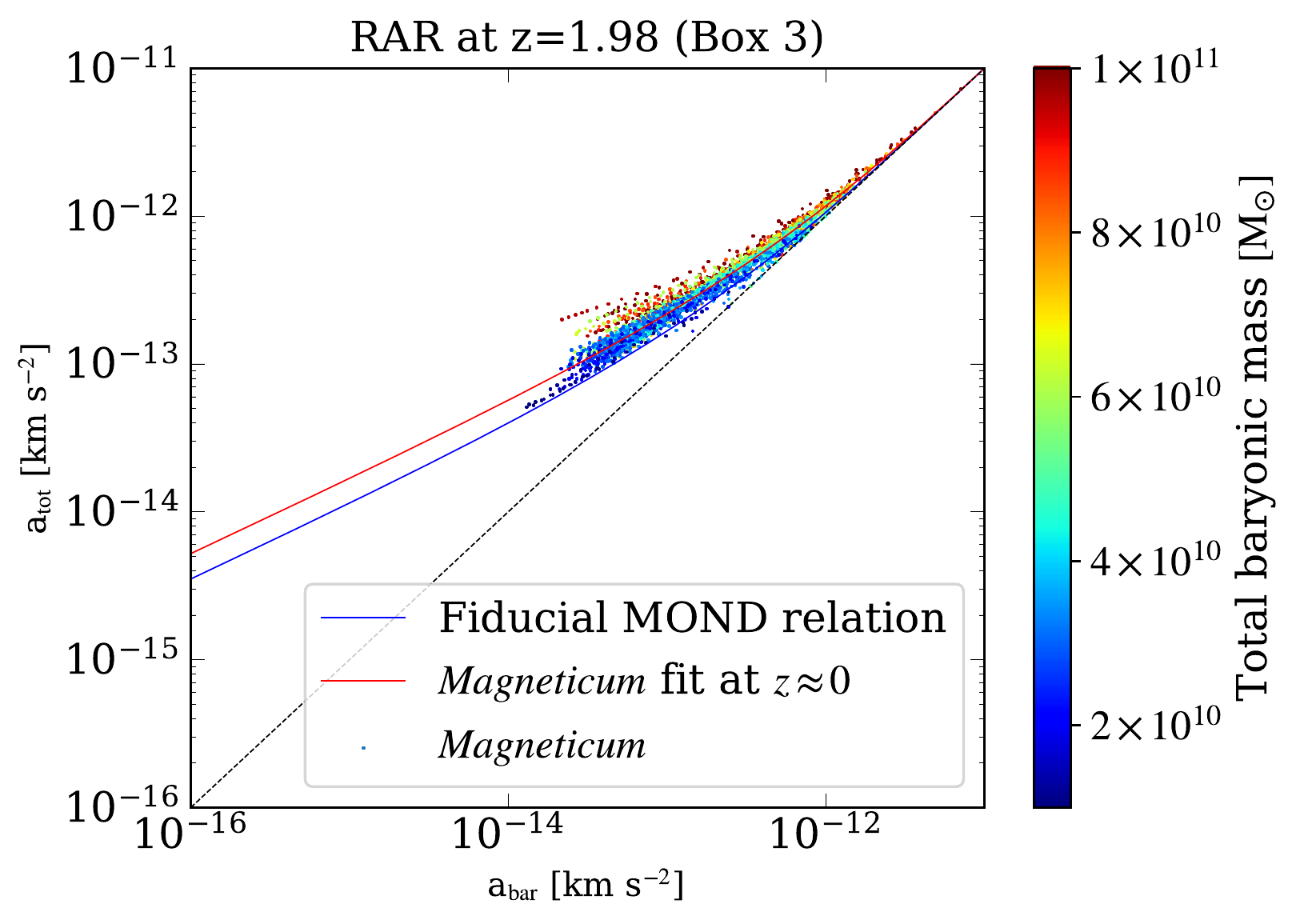}
\caption{MDAR and RAR as before, but for the larger galaxy sample extracted from \textit{Box3/uhr} at $z\approx 2$.}
\label{biggerbox}
\end{figure}

Figure \ref{biggerbox} displays the MDAR and RAR for the disk galaxies extracted from the larger volume simulation (e.g. \textit{Box3/uhr} of the \textit{Magneticum} set. As mentioned before, the best-fit values for $a_0$ shows only a negligible change from the one obtained from the smaller galaxy sample extracted from \textit{Box4/uhr} and is also substantially higher than the one obtained at $z\approx 0$ in that box.

\bsp	
\label{lastpage}
\end{document}